\documentclass[
  journal=pasa,
  manuscript=research-paper, 
  year=2021,
  volume=xx,
]{cup-journal}

\usepackage{microtype,siunitx,booktabs}
\usepackage{bm}
\usepackage{graphicx}	
\usepackage{amsmath}	
\usepackage{amssymb}	
\usepackage[font=small,labelfont=bf]{caption}
\title{System design and calibration of SITARA - a global 21~cm short spacing interferometer prototype
}

\author{Jishnu N. Thekkeppattu}
\affiliation{International Centre for Radio Astronomy Research, Curtin University, Bentley, WA 6102, Australia}
\alsoaffiliation{Raman Research Institute, C V Raman Avenue, Sadashivanagar, Bengaluru, 560080, India}
\email[J.N. Thekkeppattu]{j.thekkeppattu@postgrad.curtin.edu.au}

\author{Benjamin McKinley}
\affiliation{International Centre for Radio Astronomy Research, Curtin University, Bentley, WA 6102, Australia}
\alsoaffiliation{ ARC Centre of Excellence for All Sky Astrophysics in 3 Dimensions (ASTRO 3D), Bentley, WA 6102, Australia}

\author{Cathryn M. Trott}
\affiliation{International Centre for Radio Astronomy Research, Curtin University, Bentley, WA 6102, Australia}
\alsoaffiliation{ ARC Centre of Excellence for All Sky Astrophysics in 3 Dimensions (ASTRO 3D), Bentley, WA 6102, Australia}

\author{Jake Jones}
\affiliation{International Centre for Radio Astronomy Research, Curtin University, Bentley, WA 6102, Australia}

\author{Daniel C. X. Ung}
\affiliation{International Centre for Radio Astronomy Research, Curtin University, Bentley, WA 6102, Australia}

\doi{10.1017/pasa.xxxx.xx}

\received {dd Mmm YYYY}
\revised  {dd Mmm YYYY}
\accepted {dd Mmm YYYY}
\published{dd Mmm YYYY}


\begin{document}

\begin{abstract}
Global 21--cm experiments require exquisitely precise calibration of the measurement systems in order to separate the weak 21--cm signal from Galactic and extragalactic foregrounds as well as instrumental systematics. Hitherto, experiments aiming to make this measurement have concentrated on measuring this signal using the single element approach. However, an alternative approach based on interferometers with short baselines is expected to alleviate some of the difficulties associated with a single element approach such as precision modelling of the receiver noise spectrum. Short spacing Interferometer Telescope probing cosmic dAwn and epoch of ReionisAtion (SITARA) is a short spacing interferometer deployed at the Murchison Radio-astronomy Observatory (MRO). It is intended to be a prototype or a test-bed to gain a better understanding of interferometry at short baselines, and develop tools to perform observations and data calibration. In this paper, we provide a description of the SITARA system and its deployment at the MRO, and discuss strategies developed to calibrate SITARA. We touch upon certain systematics seen in SITARA data and their modelling. We find that SITARA has sensitivity to all sky signals as well as non-negligible noise coupling between the antennas. It is seen that the coupled receiver noise has a spectral shape that broadly matches the theoretical calculations reported in prior works. We also find that when appropriately modified antenna radiation patterns taking into account the effects of mutual coupling are used, the measured data are well modelled by the standard visibility equation.
\end{abstract}




\section{Introduction}
\label{sec:Intro}

The period in cosmological history when the first stars ionised the Universe remains one of the least constrained epochs in the concurrent cosmological models. This period known as the cosmic dawn and epoch of reionisation (CD/EoR), despite being a critical epoch in our cosmological models, lacks observational constraints. It has been recognized that the redshifted signal from the 21~cm hyperfine transition of neutral hydrogen can be an effective tracer of baryonic evolution during this period \citep{1977SvAL....3..155V, PhysRevD.82.023006}. The coupling of this transition's spin temperature to radiation temperatures (CMB as well as any excess background radiation) via scattering, and matter kinetic temperature via collisions as well as the Wouthuysen-Field effect \citep{1952AJ.....57R..31W, 1958PIRE...46..240F} can give rise to absorption and emission features in the mean background spectrum. Owing to cosmological expansion, the rest-frame frequency of 1420~MHz of this transition gets redshifted to 40--230~MHz. There is considerable effort being put in to measure the spatial power spectrum of this signal, with several radio telescopes such as MWA \citep{2013PASA...30....7T, 10.1093/mnras/staa414}, LOFAR \citep{2013A&A...556A...2V, 10.1093/mnras/staa327}, GMRT \citep{1991CuSc...60...95S, 10.1111/j.1365-2966.2011.18208.x}, HERA \citep{DeBoer_2017, 2021arXiv210802263T}, 21 centimeter Array \citep{2004MPLA...19.1001P}, OVRO-LWA \citep{2015AAS...22532801H, 10.1093/mnras/stab1671} currently operating with precision measurements of 21~cm power spectrum as one of the key science goals. The sky-averaged or global component has also been recognized as a powerful probe of this epoch \citep{1999A&A...345..380S}. Since this uniform component is an average of the angular variations, a single antenna of low angular resolution is sufficient to detect the signal. Given that the sky-averaged component has a strength of $\rm \sim 10-100~mK$ against Galactic and extragalactic foregrounds with $\rm 10^2-10^4~K$ brightness temperatures, an unambiguous detection of this signal requires well calibrated instruments. Most of the experiments aiming at a measurement of the global signal, such as EDGES \citep{Bowman2018}, SARAS \citep{2018ExA....45..269S, T.2021}, BIGHORNS \citep{2015PASA...32....4S}, PRIZM \citep{2019JAI.....850004P}, LEDA \citep{doi:10.1093/mnras/sty1244}, REACH, MIST, HYPEREION (Patra et.al., in prep) use single well-calibrated antennas as the electromagnetic sensor. However, these experiments require precision calibration of the systems to mitigate the effects of the antenna transfer function, antenna radiation pattern variations with frequency (beam chromaticity) as well as the receiver bandpass and spectrum of the receiver noise. 

\section{Background and motivation}
\label{sec:background}
As an alternative to single antenna based measurement of the 21~cm signal, interferometers with closely spaced antennas have been proposed. The motivation for interferometers stems from the fact that individual receiver noise contributions, being uncorrelated, average to zero upon cross-correlation. Conventional wisdom based on a Fourier perspective is that an interferometer does not respond to a uniform sky signal. However, this argument fails at the limit when the antennas are brought to close proximity. 

A radio interferometer measures the spatial coherence function. For wavelength $\lambda$ corresponding to a frequency $\nu$ and for a baseline vector $\Vec{b}$, the coherence is given by Eq.\ref{Eq:coh_1} (see for e.g. \cite{1999ASPC..180....1C}), 
\begin{equation}
    \label{Eq:coh_1}
    V = \frac{1}{4\pi}\int_{4\pi} T_{sky} A_a e^{-2\pi i (\frac{\Vec{b}.\Vec{r}}{\lambda})} d\Omega
\end{equation}
where $c$ is the speed of light, $T_{sky}$ is the sky brightness temperature as a function of spatial coordinates, $A_a$ is the antenna radiation pattern, assumed to be identical for both antennas. From this, we can compute the expected auto-correlation powers for the individual antennas as well as their cross power by appropriately setting $\Vec{b}$. Setting $|b|=0$, the auto-correlation powers may be recovered. Our interest is when $|b| \sim \lambda$, where Eq.\ref{Eq:coh_1} yields a non-negligible non-zero value.

Indeed it is shown in \cite{10.1093/mnras/stv746} using a spherical harmonic expansion (instead of a Fourier expansion) that the interferometer response to a global signal has a characteristic $sinc$ shape as a function of baseline length. There also appears to be some controversy regarding the nature of this response. While \cite{Presley_2015} argue that the response is due to the primary radiation pattern of the antennas, \cite{Singh_2015} demonstrate using simulations with isotropic antennas that the response is an inherent property of the wavefield as opposed to being purely an instrumental response. Also in \cite{Singh_2015}, simulations of the coherence function as a function of baseline length $|\Vec{b}|$ for various types of antennas and orientations are shown. However, these studies ignore effects such as antenna mutual coupling, noise coupling between antennas, ground, and foregrounds.

While Eq.\ref{Eq:coh_1} provides a convenient starting point for short-spacing interferometry, it assumes identical radiation patterns for the antennas - a condition that is not necessarily satisfied due to mutual coupling when antennas are closely spaced. An interesting theoretical discussion on the effects of mutual coupling on the response of a short-spacing interferometer - from the perspective of the incomplete nature of Eq.\ref{Eq:coh_1} - is given in \cite{Venumadhav_2016}. Specifically, the "shadowing" of antennas when closely spaced and mutually coupled is not considered by Eq.\ref{Eq:coh_1}. Therefore, \cite{Venumadhav_2016} show that effects such as scattering and shadowing have to be included. It is argued in \cite{Venumadhav_2016} that cross-talk between the antennas forming a short-spacing interferometer is crucial to having a response to the sky monopole, as shadowing effects obstruct the view of antennas to regions of sky that dominate the nonzero response. In the same work, it is shown that the sensitivity of closely packed antenna arrays to a sky monopole maximises in the regime where antennas couple by non-radiative fields. However, cross-talk can also result in noise coupling between the antennas, thereby invalidating the assumption of negligible noise bias in cross-correlations.

Though there have been theoretical and simulation studies on the short baseline response of an interferometer to an all sky component of the sky, only a few experiments have attempted a measurement. ZEBRA \citep{6051278} used a resistive spatial beamsplitter made out of discrete resistors to enhance the short spacing response \citep{7265020}. While a beamsplitter enhances the coherence of a uniform sky signal at short baselines, modelling emission from the splitter, which appears as an additive term in the spectrum, presents a formidable challenge. In \cite{10.1093/mnras/staa2804}, an alternative approach based on the Engineering Development Array (EDA-2) deployed at the Murchison Radio-astronomy Observatory (MRO) has been employed to evaluate the potential of this idea. However, the presence of a large number of antennas in close proximity introduces complicated mutual coupling responses between the antennas, the effects of which are in general hard to characterise in-situ. To the best of the authors' knowledge no such study has been undertaken in literature wherein the nature of short baseline interferometer response to an all sky signal has been experimentally investigated with a dedicated experiment.

In this context, it was recognized that a dedicated broadband interferometer to study the effects of mutual coupling, noise coupling and foregrounds, and their effects in an interferometer for probing the global 21~cm signal is required.  Short spacing Interferometer Telescope probing cosmic dAwn and epoch of ReionisAtion - SITARA is the first in a series of experiments aiming at measurement and validation of the short spacing interferometer response with an ultimate aim of having a dedicated interferometric array with multiple short baselines, named All-Sky SignAl Short-Spacing INterferometer (ASSASSIN). As a first step in this direction we built and deployed a prototype two-element broadband interferometer at the MRO, to measure the response of an interferometer to the radio sky at short baselines ($\sim$ $\lambda$). This version of the instrument is envisaged to be a test-bed to develop techniques for system design, calibration and data analysis at short baselines and to understand potential systematics. Experience gained from this version will feed into more advanced experiments. In this paper we outline the SITARA system concept, deployment and data calibration strategies with a particular emphasis on the calibration of short-baseline interferometric data.

In closely spaced interferometers such as SITARA, cross-talk between the antennas becomes non-negligible. The term cross-talk can imply a wide range of phenomena; however in this paper we use cross-talk as a blanket term for any coupling of signals from one arm of the interferometer to the other. Cross-talk can occur at multiple points in the signal chain. However for short-spacing interferometers, the dominant form of cross-talk is expected to be due to the mutual impedance between the antennas. There are two major effects expected due to such cross-talk.
\begin{enumerate}
    \item The receiver noise from an antenna and associated electronics leaks into the other antenna. This results in a non-zero cross-correlation between receiver noises. This appears as a constant excess receiver noise in cross-correlations.
    \item Similarly, sky signals will also get coupled between the antennas. This results in the autocorrelations and cross-correlations deviating from the idealised simulations using the visibility equation, even if accurate antenna radiation patterns are used.
\end{enumerate}
In this paper, we find that both effects are seen in SITARA data. We also find that ignoring the effects of cross-talk leads to poor modelling of data and therefore a model for cross-talk is presented that captures the complexity of the data. 

\subsection{Notations and conventions}
\label{subsec:notation}
In this paper, we use boldface letters to denote matrices. Vectors are denoted with an arrow over the symbol, such as $\Vec{b}$. The forward Fourier transform carries a negative sign. The imaginary number is denoted by \textit{i}. We use $T$ to denote temperature quantities that are expressed in kelvins (K), and powers that have arbitrary units due to scaling are denoted as $P$. Visibilities are represented by $V$ and electric fields with $\Vec{E}$. Frequency is denoted by $\nu$ and voltage by $e$; it may be noted the difference between the use of $e$ as a voltage and as the exponential factor will be self-evident from the context. 

\section{SITARA System overview}
\begin{figure}
	\includegraphics[width=1.0\columnwidth]{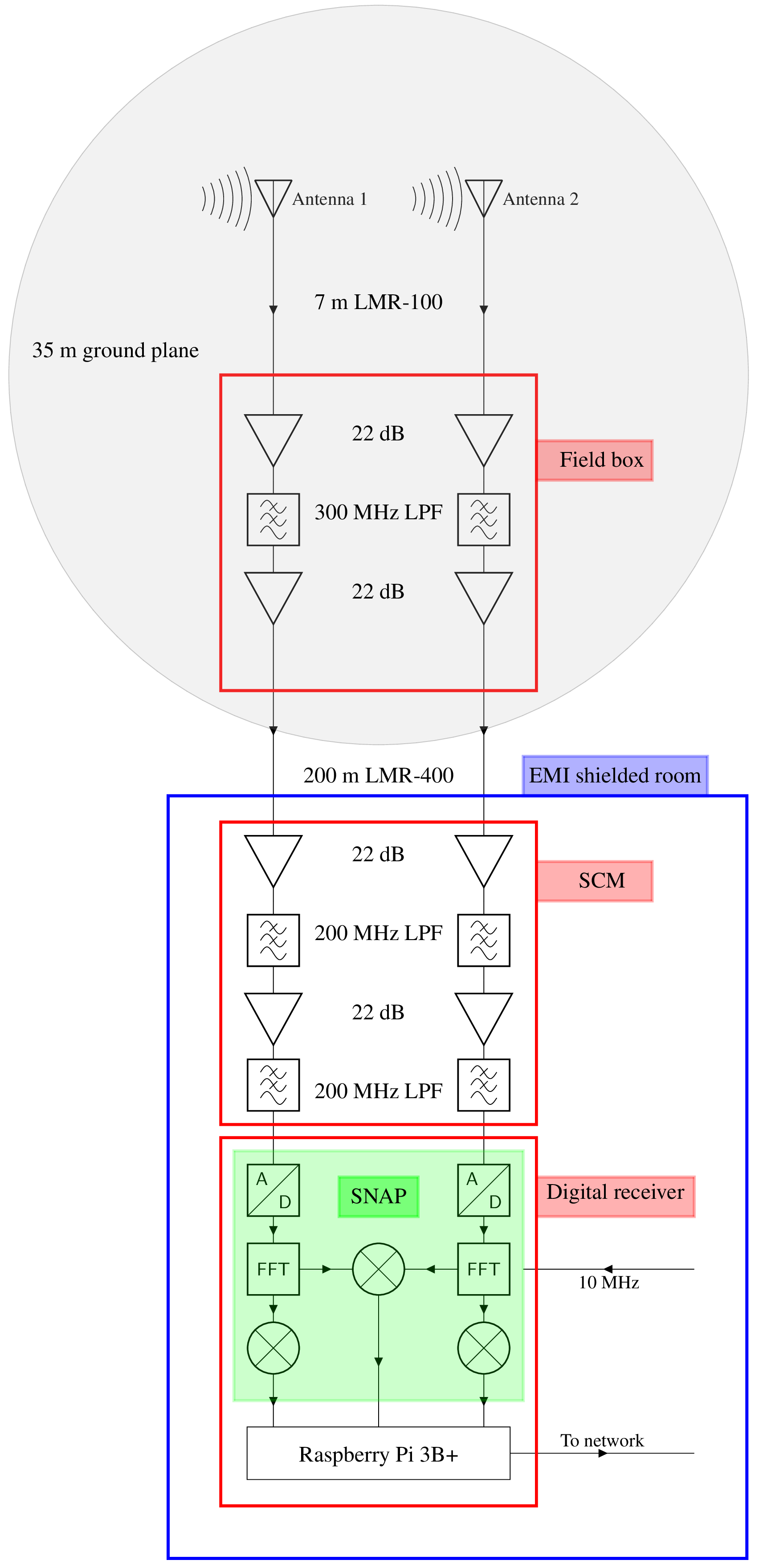}
    \caption{A high level block diagram of SITARA; auxiliary details such as power supplies as well as attenuators used for impedance matching between various modules are not shown. The multiplication units shown in the digital receiver perform conjugate multiplication.}
    \label{fig:Block_dia}
\end{figure}

Broadly, the SITARA system consists of two antennas kept in close proximity, a "fieldbox" performing initial analog signal conditioning and a back-end performing further analog signal conditioning, digitisation and correlation. No form of hardware calibration such as noise diodes is employed. The first prototype is kept simple so as to study systematics that have to be considered for more advanced designs. A block-diagram of the SITARA system is given in Fig.\ref{fig:Block_dia}

In order to enable rapid development, prototyping and deployment, it was decided to use system components with good pedigree, especially in the harsh field conditions of MRO. The two antennas used are standard MWA active dipoles kept over a metallic groundplane, each one consisting of two bow-ties forming a crossed dipole and the associated low noise amplifiers (LNA). The ground plane has a diameter of 35~m with 5~cm square grids and was previously used for the Engineering Development Array (EDA) - 1  \citep{wayth_eda_1}. Each bow-tie dipole has an end to end length of 74~cm, and a height of 40~cm. Each crossed dipole antenna, formed by two orthogonal dipoles, is held above the ground plane with four 10~cm dielectric stand-offs; one on each arm of the antenna. However, only one polarisation of each antenna is utilised for this experiment. Further details regarding the mechanical structure of the antenna can be found in \cite{Reeve_MWA}.  The central hub of each antenna houses a dual LNA assembly based on Broadcom ATF-54143 pHEMT, with a gain of about 20~dB per polarisation. The LNA circuit also performs differential to single ended conversion, such that the balanced antenna (dipole) can be connected to an unbalanced transmission line (coaxial cable). The signals from the antennas are transported over 7~m of KSR-100 coaxial cables (specification conforming to LMR-100, impedance 50~$\Omega$) to a "fieldbox" that contains modular amplifiers and filters. Through the same coaxial cables the DC power for the LNAs is supplied by the fieldbox via bias-tees.  

The fieldbox amplifies the signals further to reduce effects of a long cable on the net system temperature. To reduce effects of out-of-band radio frequency interference (RFI) on signal chain linearity (as the amplifiers used are broadband compared to the required 250~MHz bandwidth) a relatively broad 300~MHz low-pass filtering is performed. The amplified and filtered signals are transported over 200~m of coaxial cables (specification conforming to LMR-400, impedance 50~$\Omega$) to the back-end electronics housed inside a shielded room, colloquially called the Telstra hut. Power to the fieldbox is delivered over a pair of dedicated power lines running 12V DC. This power is derived from a power supply housed in the Telstra hut and passed through dedicated filters to reduce electromagnetic interference (EMI) and meet the radiated EMI specification requirements of the MRO. 

The signals arriving at the Telstra hut end are further amplified to ensure a sufficiently high signal to noise ratio to overcome the quantisation noise of the analog to digital converters (ADC) in the correlator. Two stages of 200~MHz filtering are utilised as an anti-aliasing filter to limit the bandpass to 250~MHz. Altogether, the analog section has a net gain of about 70~dB inclusive of the cables and active antenna LNA. A SNAP board \citep{doi:10.1142/S2251171716410014} sampling at 500~MSPS is used as the digitiser and correlator. Though SNAP has 12 inputs,  only 6 inputs can be utilised at the sampling rate of 500~MSPS. However this is not a constraint, as only two of the six available inputs are used for the current experiment. To reduce the amount of correlated board noise, two physically different ADCs out of the three on board are used to digitise the data. The 10~MHz clock to SNAP is provided by a White Rabbit\footnote{\url{https://ohwr.org/projects/white-rabbit}} unit referenced to a master hydrogen maser. The data are channelised into 8192 channels and correlated in the SNAP to form auto and cross-correlation spectral products. As the sampling is real, only 4096 channels are useful. Thus, each of the resulting correlated spectra has 4096 channels spanning a frequency range of 0-250~MHz, with a spectral resolution of $\rm \approx 61~kHz$. The amplifiers in the signal chain have high pass filtering at 50~MHz and the LNAs in the active antennas have high pass filtering at 70~MHz. Therefore, frequencies between 50 and 70~MHz have reduced sensitivity. The anti-aliasing filters in the analog signal chain reduce sensitivity above 200~MHz. Thus, owing to the filtering introduced by the active antennas and the analog signal processing, only frequencies between 50 and 200~MHz have sensitivity to sky signals.  

When operated at 500~MSPS, the SNAP ADCs perform interleaved sampling. Small offset, gain and phase (OGP) mismatches between the ADC cores introduce spurious tones in the spectra at sub-harmonics of the clock signal. While these tones themselves are not deleterious, as they can be flagged during data analysis, the interleaving process has been found to cause intermodulation products in the measured spectra. In principle it is possible to measure and correct for the mismatches using bin-centred tones at each frequency, however any such correction would have to be performed in the signal processing within FPGA. As the complexity required to introduce such tones without causing conducted EMI outweighs any advantage obtained, we do not perform it. 

A Raspberry Pi 3B+ (henceforth RPi) with 32~GB microSD storage, connected over GPIO controls the SNAP. The same connection is utilised to transfer correlated data to the RPi as well as provide DC power to it. An acquisition code in the RPi acquires auto and cross data from SNAP and writes them out in \texttt{miriad} format \citep{1995ASPC...77..433S} into the RPi microSD card, at a time cadence of about 3 seconds. These data are also appropriately time (UTC) and local sidereal time (LST) stamped for subsequent analysis. To reduce the EMI generated by the SNAP digital clocks from getting radiated and conducted via power lines, a dedicated switching mode power supply (SMPS) along with input EMI filters is enclosed within the correlator rack chassis. To further reduce EMI from the correlator, an off-the-shelf media converter is enclosed that converts the electrical ethernet connection from the RPi into an optical fibre connection. Thus, this unit forms a low-EMI networked correlator that can be accessed over internet.  

Efforts have been made to keep the analog signal chains symmetric in their amplitude and phase responses, nonetheless there could be an excess delay between the arms due to component tolerances. The effect of an excess delay is to decorrelate the signals, however even for an excess path length of the order of a few metres, decorrelation is expected to be minimal as the signals are fine channelised to a resolution of $\rm \sim 61~kHz$ and correlated.

\section{Deployment and observations}
\begin{figure*}[hbt!]
\centering
\includegraphics[width=0.46\textwidth]{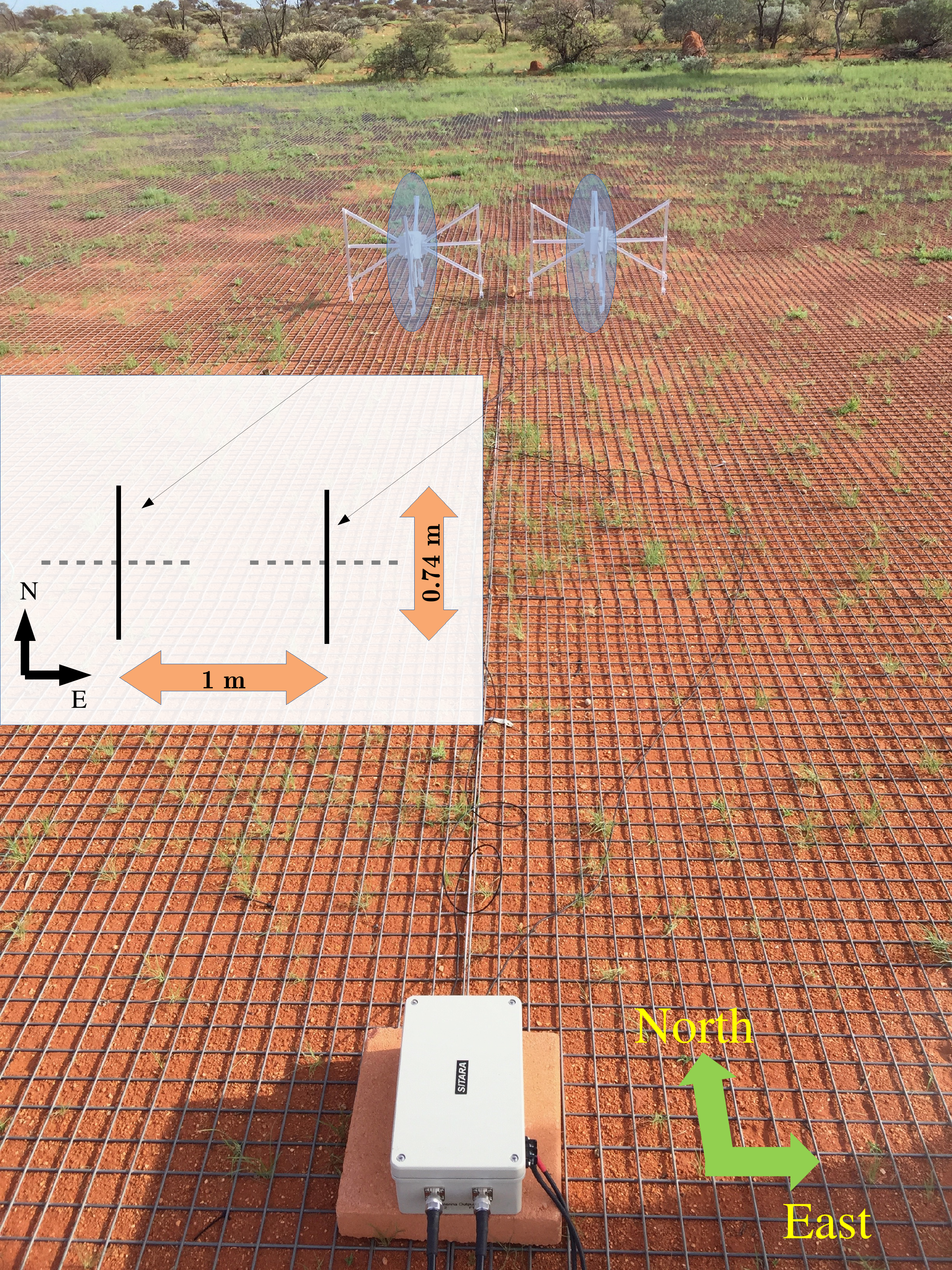}
\includegraphics[width=0.525\textwidth]{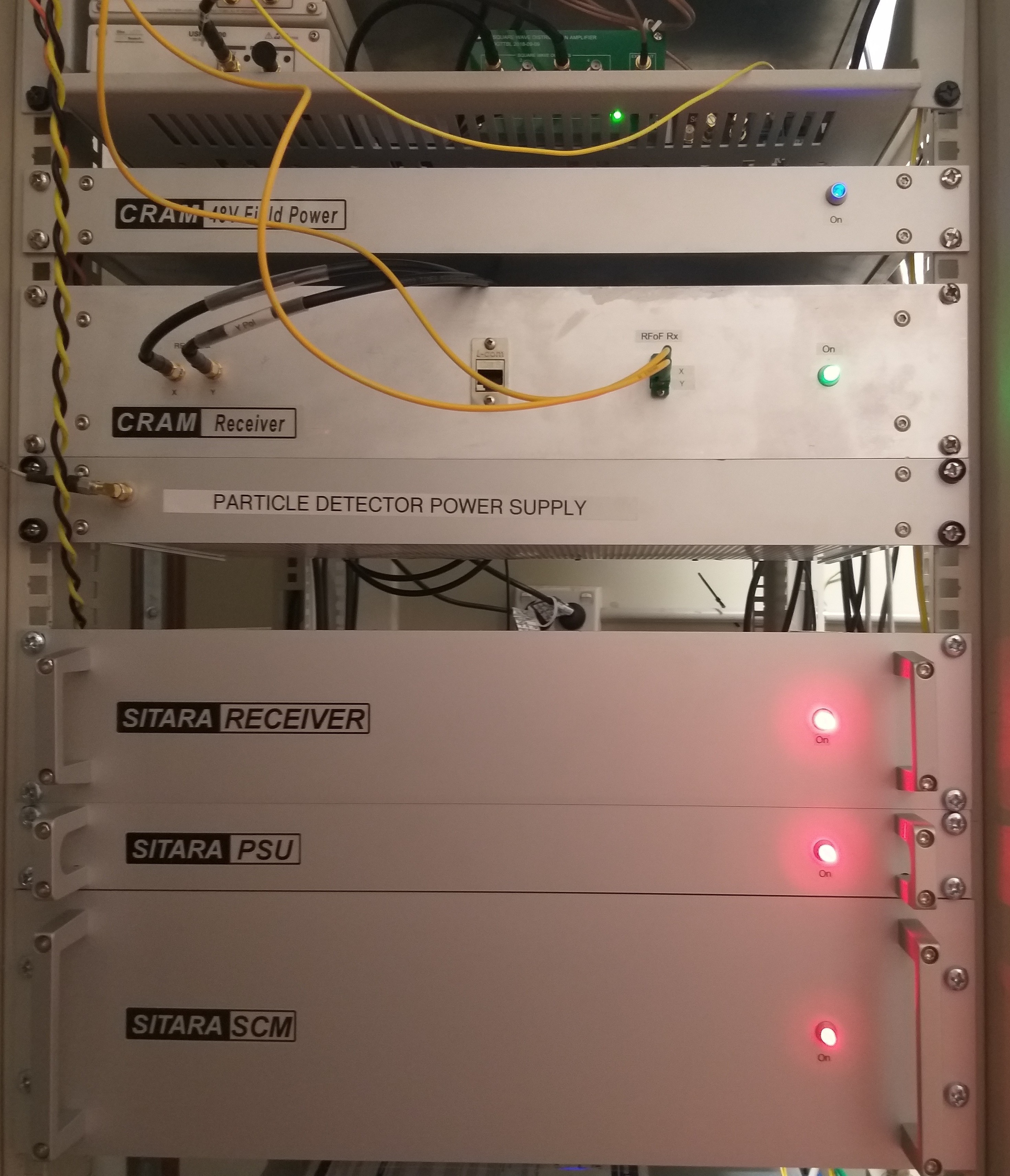}
\caption{SITARA system as deployed at MRO. The left photograph shows SITARA antennas and fieldbox; the cables have since been tied to the ground plane. The specific dipoles used in this experiment are highlighted in blue ellipses. The inset shows the antenna orientation and relevant dimensions where the inactive dipoles have been greyed out. The right photograph shows SITARA back-end electronics inside the Telstra hut. The receiver box houses the SNAP and RPi as well as media converters for networking. Signal conditioning module (SCM) contains the amplifiers and filters to perform analog processing before digitization and correlation.}
\label{fig:SITARA_MRO_figs}
\end{figure*}

SITARA was deployed at the MRO in March 2021 with first light achieved on March 10\textsuperscript{th}. The antenna spacing has been chosen as 1~m between the dipole centres, with the dipoles oriented parallel to each other - the so-called parallel configuration in \cite{Singh_2015} - along local East-West. In this configuration, the baseline is oriented along the local E-W, while the specific dipoles used have their nulls oriented along N-S.  However, after one night of observations an amplifier in one of the signal chains failed on March 11\textsuperscript{th} and had to be removed. To preserve symmetry, the corresponding amplifier in the other signal chain was also removed and brought back for further investigation of the failure. This resulted in a gain reduction of about 22~dB and the receiver temperature increased by a factor of about 2. Though the data collected before amplifier failure on March 10\textsuperscript{th} have low levels of RFI, they do not have sufficient time coverage so as to enable calibration and hence are not used in this paper. Thus, in this paper a 24 hour span of data collected after the amplifier removal are presented. 
A few photographs of SITARA as deployed at the MRO are shown in Fig.\ref{fig:SITARA_MRO_figs}.

\begin{figure*}[hbt!]
\centering
\includegraphics[width=\textwidth]{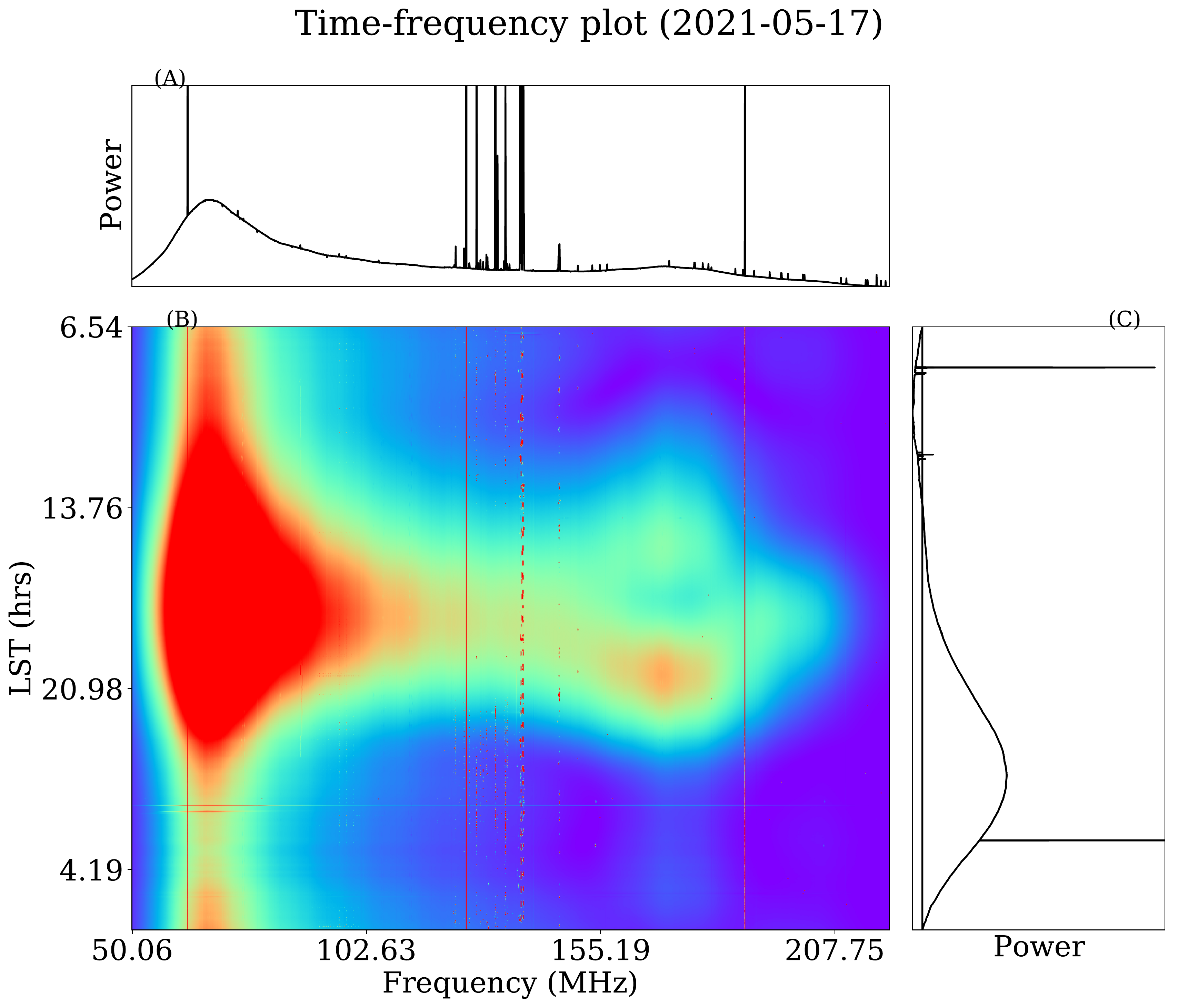}
    \caption{Time-frequency (waterfall) plot for the data collected on May 17\textsuperscript{th}-18\textsuperscript{th}, 2021. Panel B is the time-frequency plot of the magnitude of the complex cross-correlations. Panel A is the average spectrum and panel C shows the power as a function of LST for a frequency of 70~MHz. The data are unflagged and uncalibrated. The waterfall plot shows the sky drifting through SITARA beam; the peak occurs when the Galactic plane is at the local zenith. On closer inspection, the data shown in this figure are seen to contain Solar bursts between 1-2 hours LST.}
    \label{fig:waterfall}
\end{figure*}

Fig.\ref{fig:waterfall} is a time-frequency plot of the cross-correlation data collected on May 17\textsuperscript{th}-18\textsuperscript{th}, 2021. Also shown are the averages of the powers along time and frequency axes. The LNAs in the active antennas have a lower cutoff of 70~MHz, and the analog anti-aliasing filters in the analog signal chains low-pass filter the data above 200~MHz. The effects of both filters are visible in the data. Also visible are tones from ADC clocking at 62.5~MHz, 125~MHz and 187.5~MHz. These tones are of narrow-band nature and are easily flagged, however owing to the reduced analog gain due to amplifier failure there could be intermodulation products due to these tones mixing with the analog signal at frequencies close to these tones. 
The predominant sources of RFI at the MRO are satellites with downlink frequencies around 137~MHz such as Orbcomm, NOAA-APT and METEOR-LRPT weather transmissions. Amateur radio satellite downlinks around 145~MHz and aircraft communications below 130~MHz are also seen in the data. Signals from FM transmitters appear sporadically, perhaps reflected by overhead flights, meteor trails or through some VHF propagation modes such as tropospheric ducting or sporadic E-layer propagation \citep{1457406, jessop_1983}. Details about the RFI conditions at the MRO can be found in \cite{2015PASA...32....8O} and \cite{7386856}. A recent study of RFI at MRO in the broadcast FM bands is reported in \cite{tingay_sokolowski_wayth_ung_2020}.

\begin{figure*}[hbt!]
\centering
\includegraphics[width=\columnwidth]{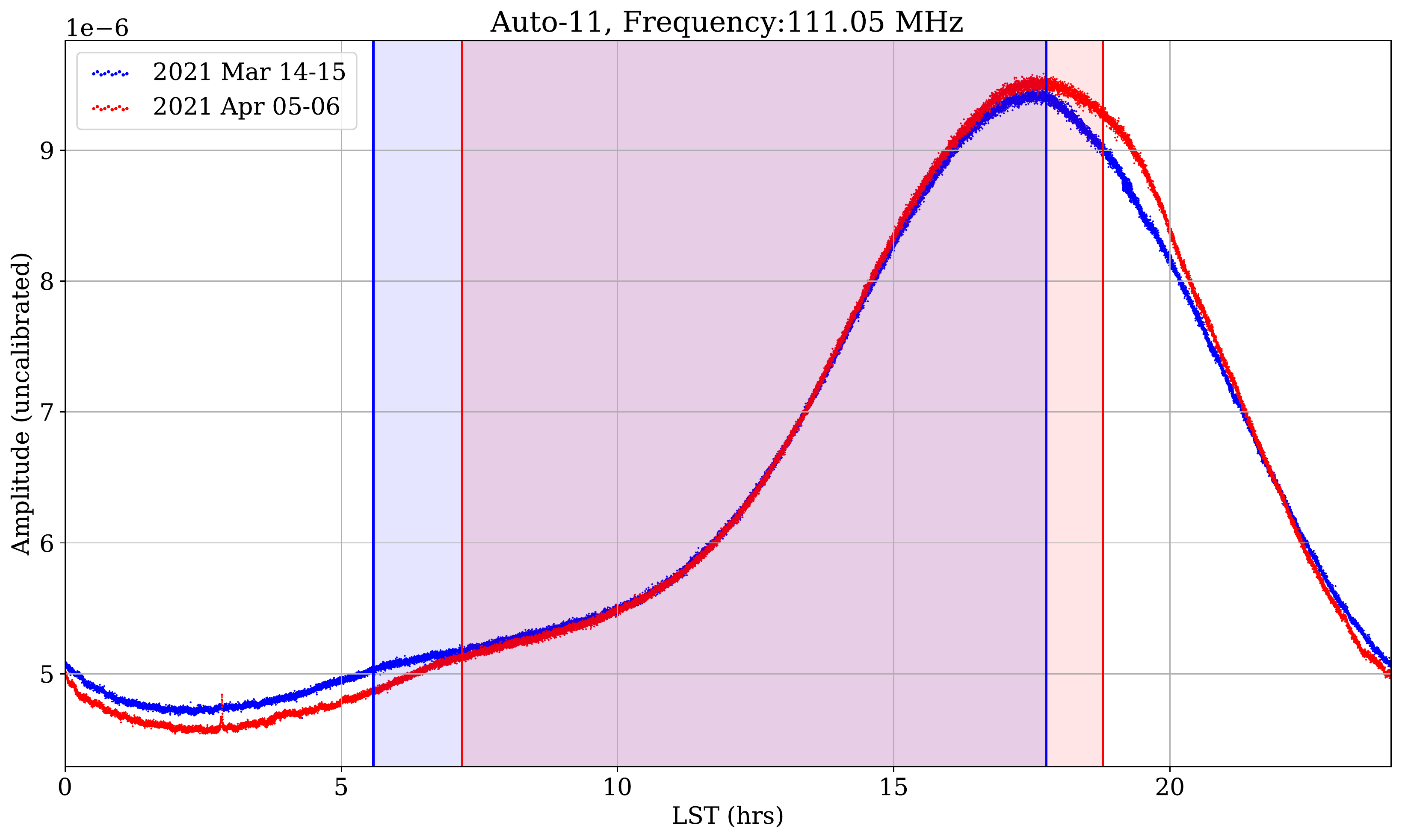}
\includegraphics[width=\columnwidth]{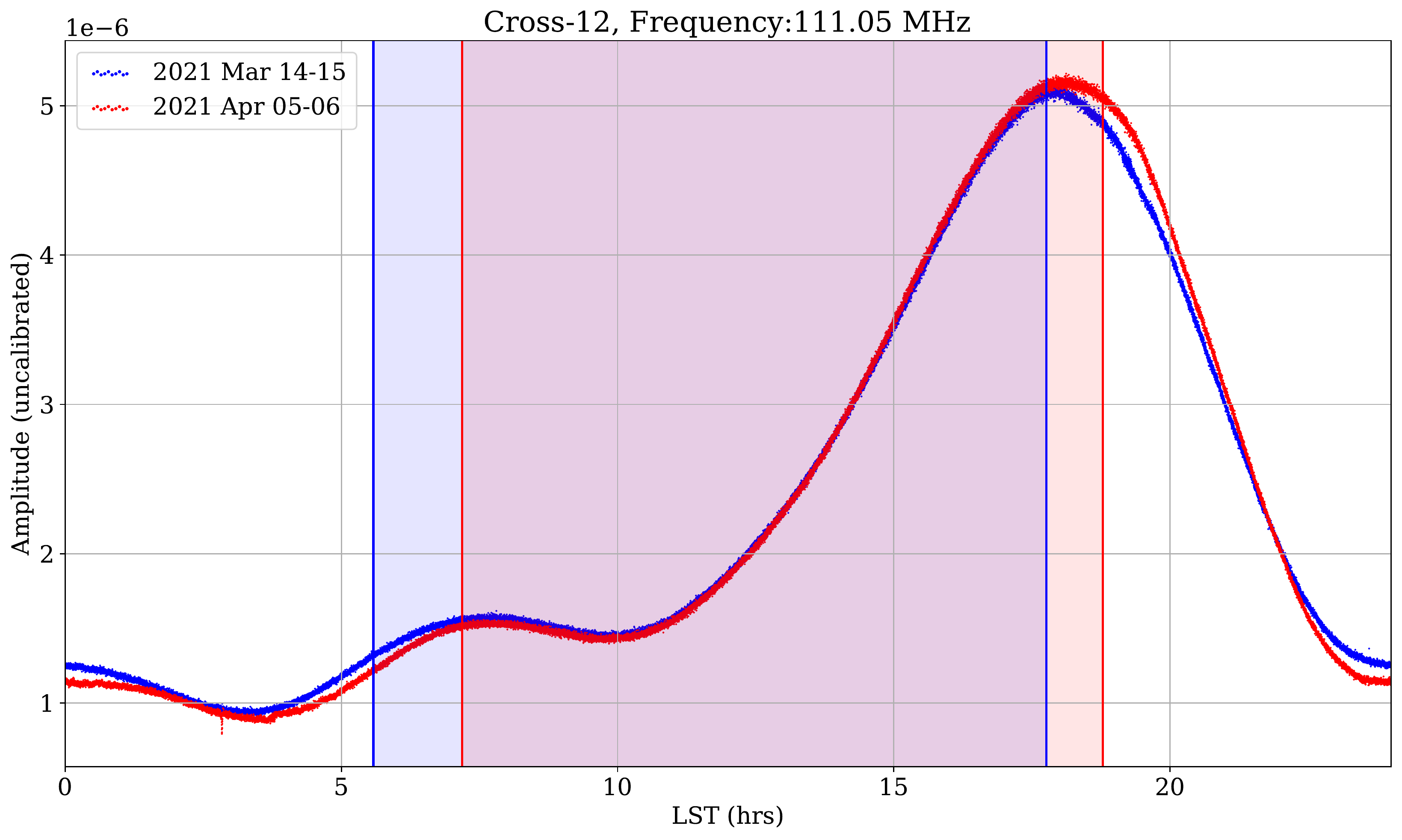}
\caption{Variations in uncalibrated power with local sidereal time (LST) for data collected on Mar 14-15, 2021 and April 05-06, 2021. The top figure shows the power in a single frequency channel in antenna 1 auto-correlations and the bottom figure shows the magnitude of antenna 1-2 cross-correlations. The colored regions in the plots show the night time LSTs for the corresponding day.}
\label{fig:cross_stab}
\end{figure*}

Data are continuously collected and the timestamped data are accessed for analysis at regular intervals. With more than 2500 hours of operation and data collection, no major glitch has been noticed. In Fig.\ref{fig:cross_stab}, the measured uncalibrated powers as a function of LST in a single 61~kHz frequency channel centred at 111.05~MHz are shown (for auto-correlation for antenna 1, as well the cross-correlation between the two antennas). The data used are after the amplifier failure. As expected, data collected with the system over a span of few weeks show variations in the power levels with time, however we do not find any significant drift with time in the system performance. 
During the ongoing observational run, we had a few serendipitous high signal to noise ratio detections of solar bursts. Analysis of those bursts are beyond the scope of this paper and will be reported elsewhere.

\section{Data calibration and analysis}
\label{sec:calib}
In this section, we describe the procedures adopted to calibrate and analyse data. The observations in \texttt{miriad} format are flagged with \texttt{pgflag} using the SumThreshold algorithm \citep{2010MNRAS.405..155O}. Further calibration and analysis of data are carried out in custom python codes with data read using the \texttt{aipy}\footnote{\url{https://github.com/HERA-Team/aipy}} package.

This section is organised as follows. Before inspecting the data, we visit the antenna radiation patterns from FEE simulations in Sec.\ref{subsec:antenna_beams}, where we find that the individual antenna patterns cannot be treated as frequency invariant. In Sec.\ref{subsec:auto_cal}, a simple model for measurements that considers cross-talk for (internal) receiver noise but not (external) sky signals is presented. We find that while this simplistic model is able to represent the variations in data, certain shortcomings are evident. The differences seen between the mock data and SITARA data are attributed to the cross-talk of sky signals between the antennas and an empirical model for it is introduced in Sec.\ref{subsec:cal_crosstalk}. This model brings the coupling of receiver noises and sky signals between the antennas under the same formalism. Interestingly, effects of sky signal cross-talk become evident in the data only at frequencies where the antenna patterns differ, as the individual auto-correlations are identical when the antenna patterns are identical. As a by-product of the model, we obtain the coupled receiver noise at all frequencies. Comparing the models with and without cross-talk, we find that the empirical cross-talk model captures the variations in the data as a function of LST accurately. 

\subsection{A prelude on antenna radiation patterns}
\label{subsec:antenna_beams}
\begin{figure*}[hbt!]
\centering
\includegraphics[width=\columnwidth]{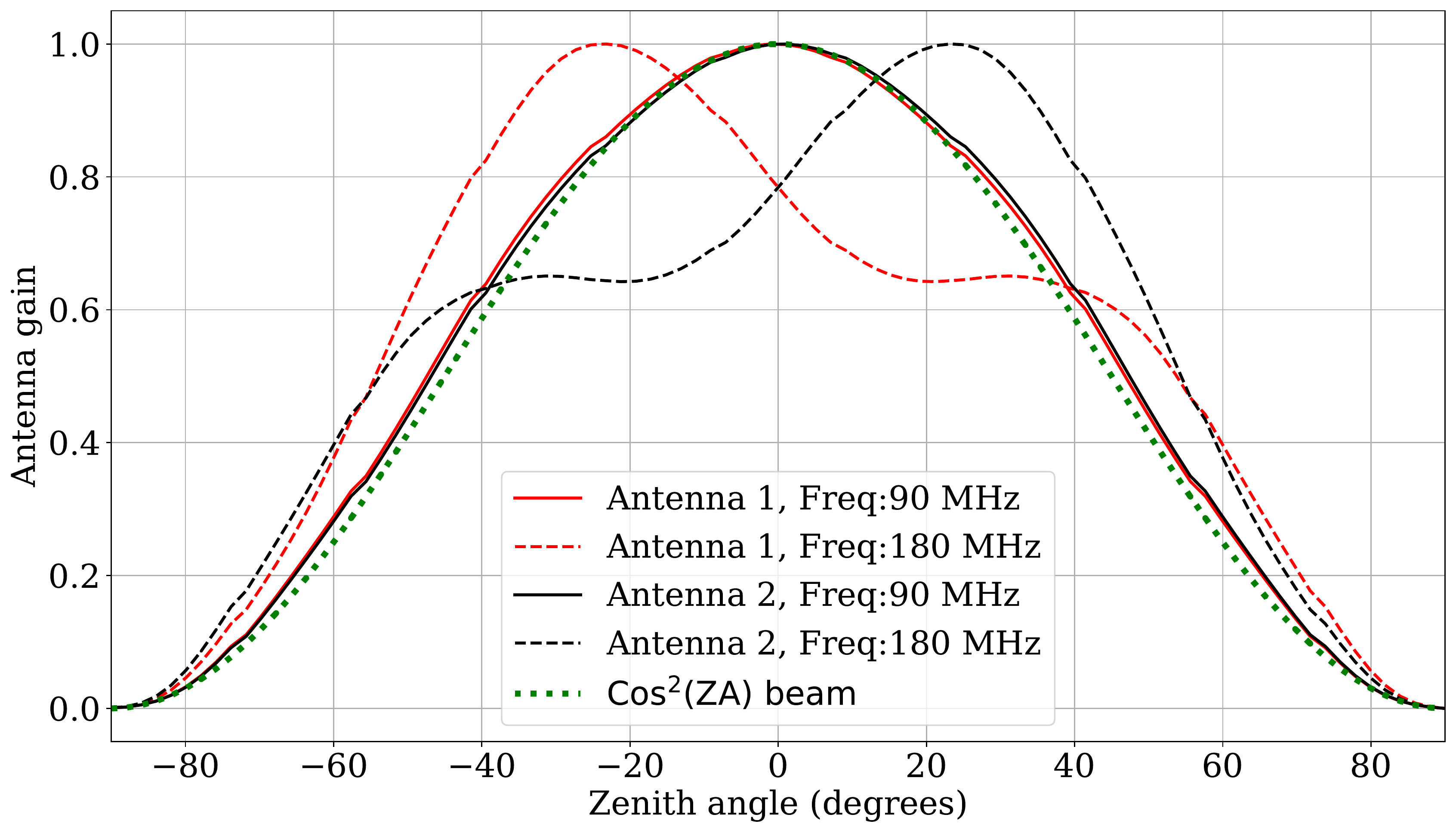}
\caption{Simulated antenna radiation patterns (H-plane) as a function of zenith angle for two MWA dipoles spaced 1~m apart in parallel configuration. The patterns at 90~MHz are identical to each other and are well approximated by an ideal dipole $\rm cos^2(ZA)$ pattern while the patterns at 180~MHz have shifted peaks away from zenith.}
\label{fig:Antenna_beam}
\end{figure*}

\begin{figure*}[hbt!]
\centering
\includegraphics[width=1.0\textwidth]{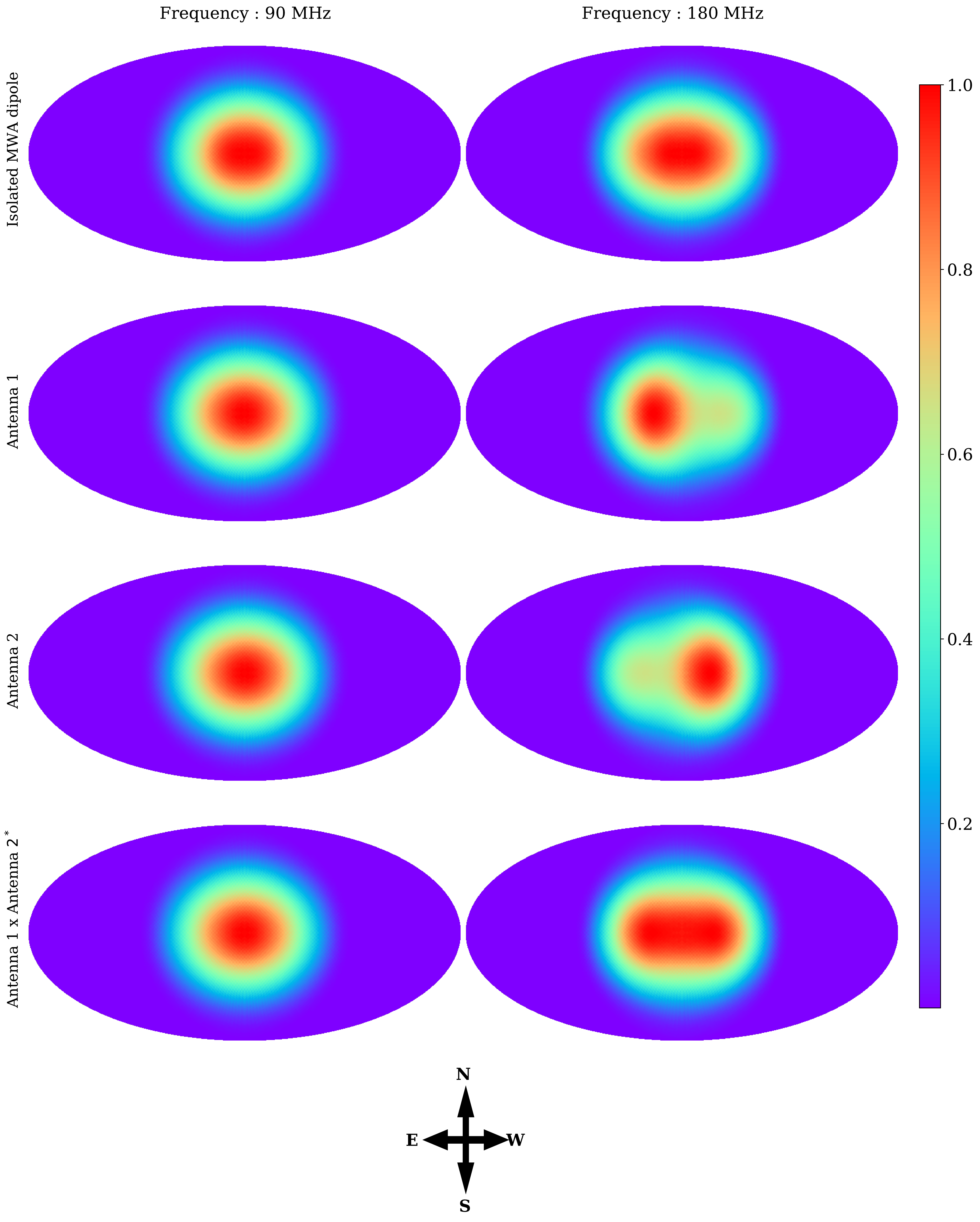}
\caption{Simulated SITARA auto and cross antenna patterns at two frequencies, in Mollweide projection. For comparison, patterns for an isolated MWA antenna are given in the top row. The plots are peak normalised as shown in the colour bar. The coordinate system is local altitude-azimuth with the centre of the Mollweide projection corresponding to zenith; the local directions are also shown. It can be seen that due to mutual coupling, the patterns of closely spaced SITARA antennas diverge from that of an isolated MWA dipole.}
\label{fig:SITARA_beams_mollview}
\end{figure*}

The radiation patterns (also called beams) of isolated MWA dipoles over a large ground plane have a peak at local zenith. With closely spaced antennas, the effects of mutual coupling between elements cause the patterns to deviate from those of isolated dipoles and hence, measured visibilities deviate from the ones computed using Eq.\ref{Eq:coh_1} if the patterns are treated as achromatic. For frequencies where the baseline $|\Vec{b}| > \lambda/2$, the antenna patterns are seen to vary rapidly with frequency with the peak shifting away from the zenith. Moreover, owing to the intrinsic symmetry, the peak shifts in opposite directions for each antenna and thus their overlapping beam solid angles vary as a function of frequency. Nonetheless, electromagnetic simulations model these and hence it is possible to use the simulated antenna patterns to compute visibilities as given in \ref{subsec:auto_cal}. For this work, we use antenna patterns simulated with FEKO \footnote{https://www.altair.com/feko/}. The simulation is for the full structure of the SITARA antenna system, which consists of two MWA dipoles at a separation of 1~m. The dipoles are assumed to be placed over an infinite ground plane. Similar to the procedure adopted in \cite{2017PASA...34...62S}, the antenna ports are loaded with lumped circuit models of the LNA. This ensures that the simulation yields the patterns of each antenna in the presence of the other, including the effects of mutual coupling, and the resulting patterns are therefore embedded element patterns (EEP).  In Fig.\ref{fig:Antenna_beam}, FEKO-simulated SITARA antenna radiation patterns at two representative frequencies of 90~MHz and 180~MHz are shown as cross-sectional plots. The patterns at 90~MHz are identical to each other and are well approximated by an ideal dipole ($\rm cos^2(ZA)$) radiation pattern, while the patterns at 180~MHz are not identical to each other. 
 A more insightful representation of the antenna patterns is given in Fig.\ref{fig:SITARA_beams_mollview}, which shows intensity maps in Mollweide projection at 90~MHz and 180~MHz of the power patterns given by \begin{equation}
    |E_{j,\theta}(\theta, \phi)E^{*}_{k, \theta}(\theta, \phi) + E_{j,\phi}(\theta, \phi)E^{*}_{k, \phi}(\theta, \phi)|
    \label{Eq:intensity_eq}
\end{equation} 
where $E_{j,\theta}$ and $ E_{j,\phi}$ are the two orthogonal components of the E-field patterns of antenna $j$; similarly $E_{k,\theta}$ and $ E_{k,\phi}$ are the components of antenna $k$. In Eq.\ref{Eq:intensity_eq}, when $j=k$, the patterns are of individual antennas, while $j \neq k$ gives the cross-correlated beam. At 90~MHz, the individual patterns are similar to that of an isolated MWA antenna, while at 180~MHz, they deviate substantially from the pattern of an MWA dipole. Moreover, the antenna patterns have a mirror symmetry owing to the inherent symmetry of a two antenna system. For subsequent analysis, we use these EEPs to simulate the expected sky response.

\subsection{Calibration ignoring sky signal cross-talk}
\label{subsec:auto_cal}
We first attempt to model the measurements with a simple model that does not take into consideration the cross-talk of sky signals between the antennas. However, excess noise temperature in cross-correlations due to cross-talk of receiver noise is considered, since neglecting it is seen to yield poor results. We model the power measured in auto-correlations and cross correlations at each frequency as affine equations as given in Eqs.\ref{eq:Temps}. 

The set of equations in Eqs.\ref{eq:Temps} is an adaptation of a commonly used system model in single element global 21~cm experiments where the measured  data are modelled as an ideal sky signal along with a multiplicative gain and an additive constant. In this model, the gains include all the multiplicative factors in the system such as the antenna efficiencies, analog gains, and any scaling introduced by the correlation and digital signal processing. The constant additive comprises of forward and reflected receiver noise and losses in the system. Similar models have been widely adopted for calibration of single element global 21~cm experiments such as EDGES \citep{https://doi.org/10.1029/2011RS004962} and SARAS \citep{T.2021}. 
\begin{align}
\label{eq:Temps}
P_{11}  &= (T_{A11} + T_{N11}) |G_1|^2  \\ \nonumber
P_{22}  &= (T_{A22} + T_{N22}) |G_2|^2  \\ \nonumber
P_{12}  &= (T_{A12} + T_{N12}) |G_1| |G_2| e^{i (\phi_1-\phi_2)} 
\end{align}
where $T_{An}$ are the respective beam-weighted sky brightness temperatures, $G_1=|G_1|e^{i\phi_1}$ and $G_2=|G_2|e^{i\phi_2}$ are the complex gains of the signal chains, $T_{N11}$ and $T_{N22}$ are the excess noise powers in the individual auto-correlations in temperature units, with the dominant contribution from the active antenna LNA. We have dropped the frequency terms for brevity. We consider the coupled receiver noise due to cross-talk in $T_{N12}$, however we ignore the cross-talk of sky signals. All temperatures are referred to the sky plane (i.e. the beam weighted sky temperature at the antenna terminals) and hence are in units of brightness temperature.

The absence of an in-situ absolute calibration, coupled with the  wide radiation patterns of the antennas, motivates a calibration with diffuse sky models. Calibration of radio telescopes using models of the diffuse sky has been successfully utilised in low frequency astronomy \citep{https://doi.org/10.1029/2003RS003016} as well as in a single antenna global 21~cm context \citep{Singh_2017}. We adopt a similar procedure to obtain the gains and use them to scale the measured data to units of kelvin (K).

To calibrate auto-correlations, expected sky spectra are computed as follows. \texttt{pyGDSM}\footnote{\url{https://github.com/telegraphic/pygdsm}}, a python implementation of the global sky model \cite[GSM;][]{2008MNRAS.388..247D, 2017MNRAS.464.3486Z}, is employed as the sky model. In this work we use the \cite{2008MNRAS.388..247D} version of GSM. For antenna radiation patterns, FEKO simulated EEPs of two MWA dipoles over a ground plane as discussed in Sec.\ref{subsec:antenna_beams} are used. As mentioned in Sec.\ref{sec:Intro}, while the visibility equation given by Eq.\ref{Eq:coh_1} is a convenient starting point, it is insufficient to compute expected sky spectra when the antennas are mutually coupled. Specifically, the assumption of identical antenna patterns for both of the antennas fails, as shown in Fig.\ref{fig:SITARA_beams_mollview}. Therefore an appropriately modified visibility equation, given by Eq.\ref{Eq:Visi_jones_a}, has to be employed.
\begin{equation}
    \label{Eq:Visi_jones_a}
    V_{jk}(\nu) = \int J_j(\nu, \hat{n}) C(\nu, \hat{n}) J_k^H(\nu, \hat{n}) e^{ (\frac{-2\pi i\nu\Vec{b}.\hat{n}}{c})} d\hat{n}
\end{equation}
where $J_j(\nu, \hat{n})$ and $ J_k(\nu, \hat{n})$ are the Jones matrices for the two antennas and $C(\nu, \hat{n})$ is the coherency matrix \citep{1996A&AS..117..137H, 2011A&A...527A.106S}. Since only a single linear polarisation is utilised in our experiment, Eq.\ref{Eq:Visi_jones_a} can be reduced into 
\begin{equation}
\label{Eq:Visi_jones_b}
\begin{split}
    V_{jk}(\nu) = \int T(\nu, \hat{n}) [E_{j,\theta}(\nu, \hat{n})E^{*}_{k, \theta}(\nu, \hat{n}) + E_{j,\phi}(\nu, \hat{n})E^{*}_{k, \phi}(\nu, \hat{n})] \\
    e^{ (\frac{-2\pi i\nu\Vec{b}.\hat{n}}{c})} d\hat{n}
\end{split}
\end{equation}
where we also simplify the coherency matrix to consist of unpolarised radiation of brightness temperature $T(\nu, \hat{n})$. Comparing Eqs.\ref{Eq:coh_1} and \ref{Eq:Visi_jones_b} we can readily see that the antenna pattern $A_a$ in Eq.\ref{Eq:coh_1}, assumed to be identical for both the antennas, can be replaced by the quantity $ A_{j,k}(\nu, \hat{n})$ given by Eq.\ref{Eq:Complex_beam}.
\begin{equation}
    \label{Eq:Complex_beam}
    A_{jk}(\nu, \hat{n}) = E_{j,\theta}(\nu, \hat{n})E^{*}_{k, \theta}(\nu, \hat{n}) + E_{j,\phi}(\nu, \hat{n})E^{*}_{k, \phi}(\nu, \hat{n}).
\end{equation}
It has to be noted that the visibilities given as per Eq.\ref{Eq:Visi_jones_b} are not normalised. All sky interferometry as performed by SITARA is greatly benefited by the use of the HEALPix \citep{2005ApJ...622..759G} framework; therefore the sky maps as well as antenna patterns are manipulated in HEALPix format. In this case, the vector $\hat{n}$ pointing to a celestial coordinate can be mapped to a pixel in a HEALPix map. With EEP simulations of antennas performed with a common origin - which is the mid point of the two antennas - the complex E-fields contain the geometrical phase referred to that common origin.  Therefore the exponential factors corresponding to geometrical phase in Eq.\ref{Eq:Visi_jones_b} can be removed, leading us to Eq.\ref{Eq:Visi_1}.
\begin{equation}
    \label{Eq:Visi_1}
    V_{jk}(\nu, LST) = \frac{\sum_{n=1}^{N_{pix}} T(n,\nu, LST)A_{jk}(n, \nu)}{\sum_{n=1}^{N_{pix}} |A_{jk}(n, \nu)|}
\end{equation}
Eq.\ref{Eq:Visi_1} is the discretized form of Eq.\ref{Eq:Visi_jones_b} where the visibilities computed are also normalised. The normalisation adopted is such that when presented with a uniform sky temperature, the autocorrelation visibilities computed as per Eq.\ref{Eq:Visi_1} have the same uniform temperature. $T(\nu, LST)$ is the $N_{pix}~X~1$ sky map at frequency $\nu$. The sky maps are rotated to bring the right ascension corresponding to the LST and the declination corresponding to the site latitude, to the zenith.

The computation in Eq.\ref{Eq:Visi_1} is repeated at a time cadence of 6 minutes and interpolated to all timestamps for which the data are collected to yield a simulated dataset similar to the measured data in time-frequency resolution, but devoid of instrumental noise and systematics. We assume that the multiplicative receiver gains and spectrum of the additive receiver noise remain constant throughout the observations and that the antenna radiation patterns are well known. Under these assumptions, the observed data along with the computed sky temperature form an overdetermined set of linear equations that may be solved to yield the system gain as well as additives with associated errors. In practice, a simple polynomial fit to simulation \textit{vs} SITARA data is performed at each frequency. It may be noted that this technique bears a resemblance to the hot-cold or Y-factor measurement commonly employed in RF noise figure measurements, however in our case, there are multiple temperature states available by virtue of sky drift. The analog electronics in the field do not have any temperature regulation, therefore temperature fluctuations can induce gain variations. Daytime data are observed to be of poorer quality due to temperature rise in the electronics, as well as ionospheric fluctuations. This, along with the fact that during this specific observation carried out in May 2021 the highest sky temperature change occurs during local night time with the Galaxy transit, prompts us to perform calibration using night time data alone.

\begin{figure*}[hbt!]
\centering
\includegraphics[width=\columnwidth]{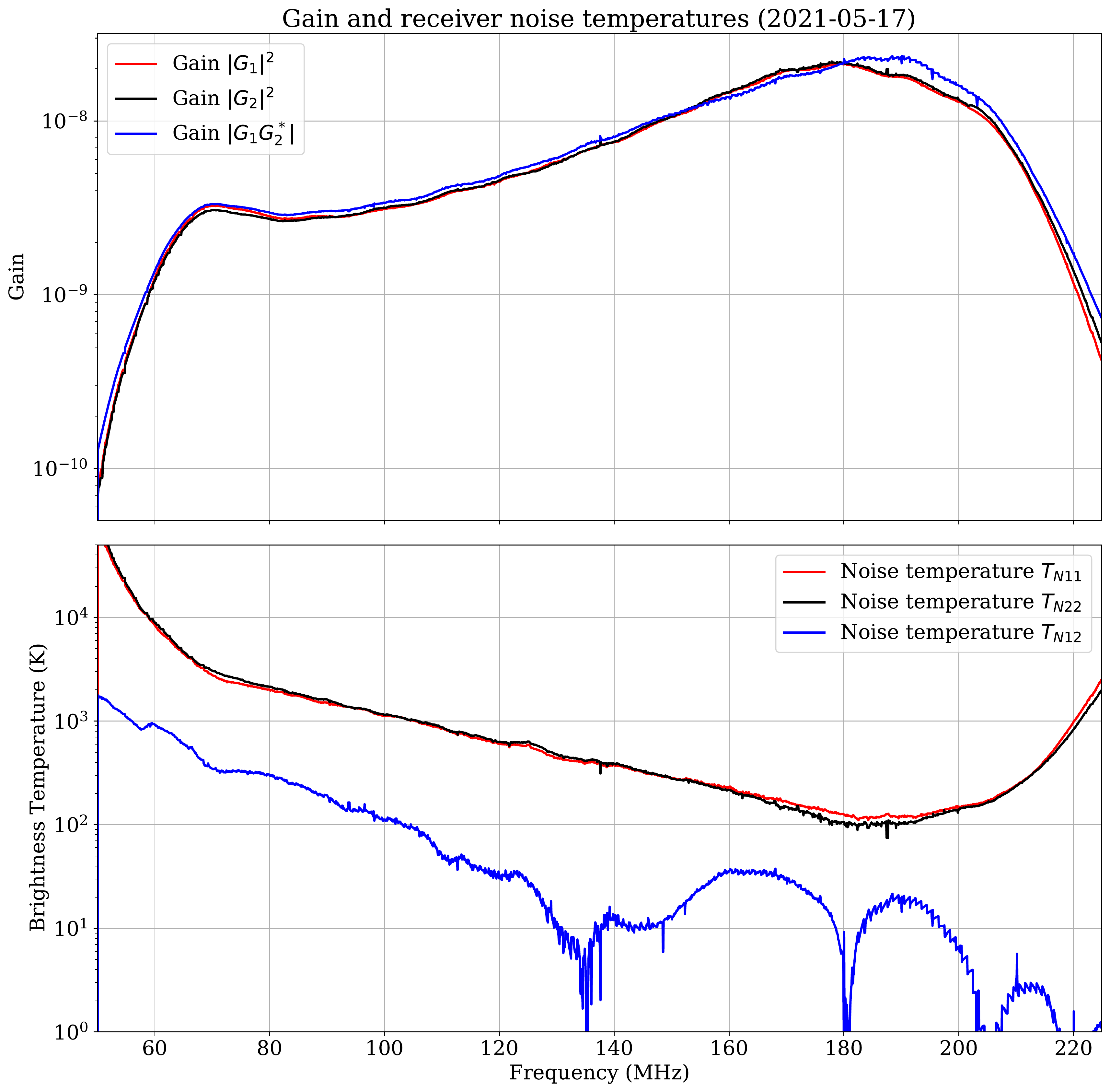}
\caption{Receiver gains and noise temperatures as functions of frequency. The plots are semi-logarithmic to accommodate a wide dynamic range. The gains show the filtering introduced by the system at 70~MHz and 200~MHz. The gains include contributions from antennas, analog stages as well as \textit{any scaling introduced by the digital signal processing in the correlator}, therefore the units are arbitrary. The noise temperatures are calibrated to units of kelvin. An interesting feature in the receiver noise temperatures is that the coupled receiver noise in cross-correlations is almost an order of magnitude less than receiver noise in autocorrelations.}
\label{fig:Gain_temp}
\end{figure*}

\begin{figure*}[hbt!]
\centering
\includegraphics[width=\columnwidth]{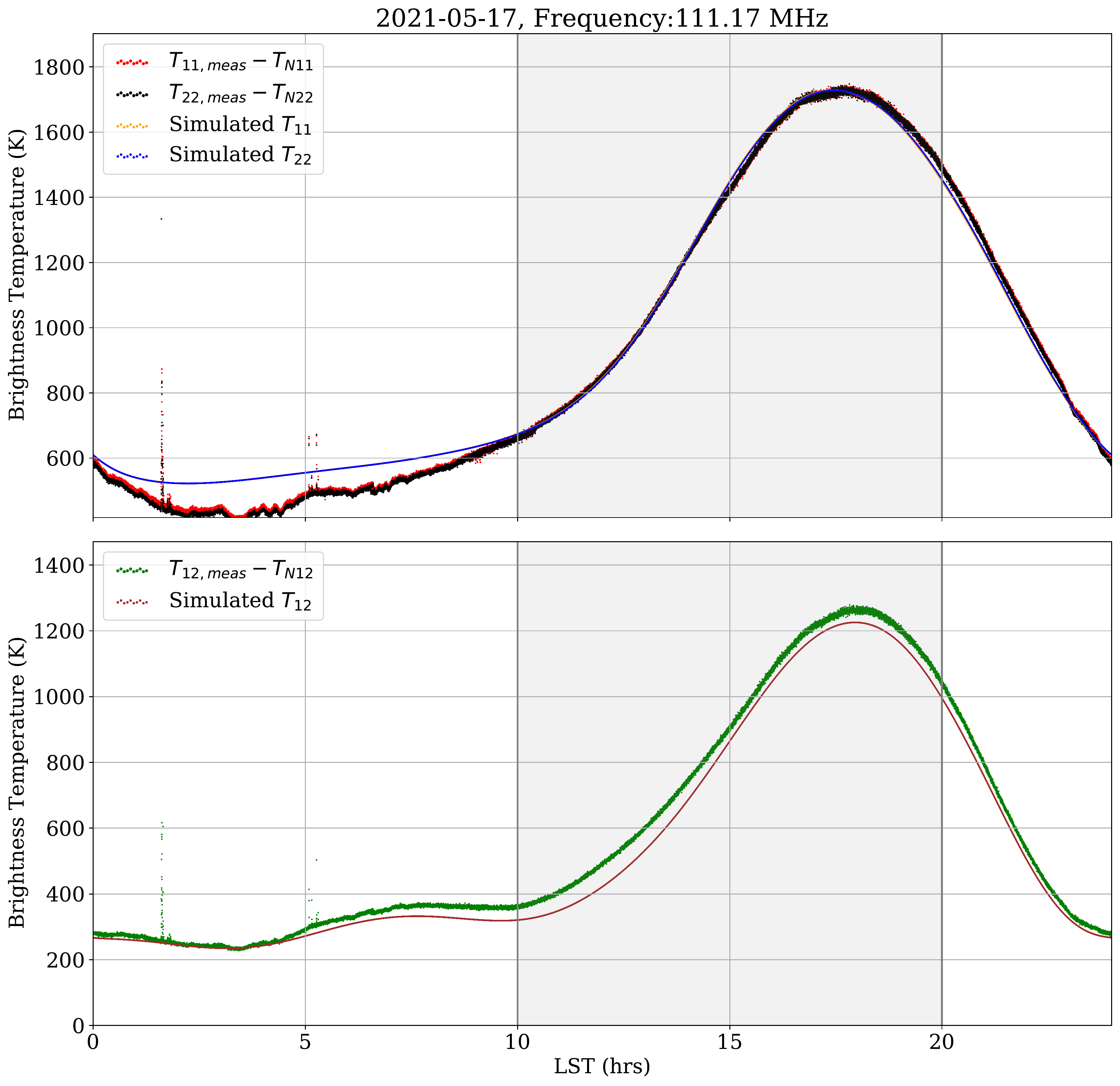}
\caption{Variations in calibrated and $T_{Nij}$ subtracted data as functions of local sidereal times (LST). The top panel shows calibrated auto-correlations along with simulated auto-correlations and the bottom panel shows calibrated cross-correlations along with simulated cross-correlations. Only data in the shaded region are used for calibration, since those LSTs have a rapid change in the sky temperature due to Galaxy transit. The solutions derived are then used for the entire data. It may be noted that $T_{Nij}$ subtraction also removes any 21~cm signal from the data.}
\label{fig:autocal_lst} 
\end{figure*}

A similar procedure is adopted to calibrate cross-correlations; however for cross-correlations the simulated and measured visibilities are complex valued. Therefore, we perform linear regression for the magnitude and phase of the visibilities separately to derive the complex gain. In addition to estimating gains, this calibration also yields an estimate of the cross-coupled receiver noise temperature. The results from calibration, viz the signal chain gains $|G_1|^2$, $|G_2|^2$ and $|G_1G_2^*|$, as well as receiver noise temperatures are shown in Fig.\ref{fig:Gain_temp}. It has to be emphasised here that the gains will have an arbitrary scaling depending on the normalisation in the FPGA firmware. Therefore the gains \textit{do not} represent the analog system gain; rather they are merely the calibration coefficients to convert observed data into units of kelvin. Also, we place less confidence on the receiver noise temperature estimates, as the model used is deemed to be incomplete since it does not take into account cross-talk between antennas. 
Estimated absolute gains $|G_1|^2$ and $|G_2|^2$ are used to calibrate the auto-correlations as shown in Eq.\ref{Eq:auto_cal_1}. 
\begin{align}    \label{Eq:auto_cal_1}
    T_{11,meas} = \frac{P_{11}}{|G_1|^2} \\ \nonumber
    T_{22,meas} = \frac{P_{22}}{|G_2|^2}
\end{align}
Similarly cross correlation visibilities are scaled with $G_1G_2^*$.
\begin{equation}
    \label{Eq:cross_cal_1}
    T_{12,meas} = \frac{P_{12}}{G_1G_2^*}
\end{equation}
Receiver noise temperatures $T_{Nij}$ may be subtracted out from $T_{ij,meas}$ for sake of comparison with simulations. Fig.\ref{fig:autocal_lst} shows the results of the calibration based on Eqs.\ref{Eq:auto_cal_1} and \ref{Eq:cross_cal_1} at a frequency of about 111~MHz. 
In Fig.\ref{fig:autocal_lst} we have also subtracted out the individual receiver noise temperatures $T_{Nij}$. We find that the simple model that we have adopted is able to capture the variations in SITARA data, at frequencies where the individual antenna patterns are somewhat identical.

\begin{figure*}[hbt!]
\centering
\includegraphics[width=\columnwidth]{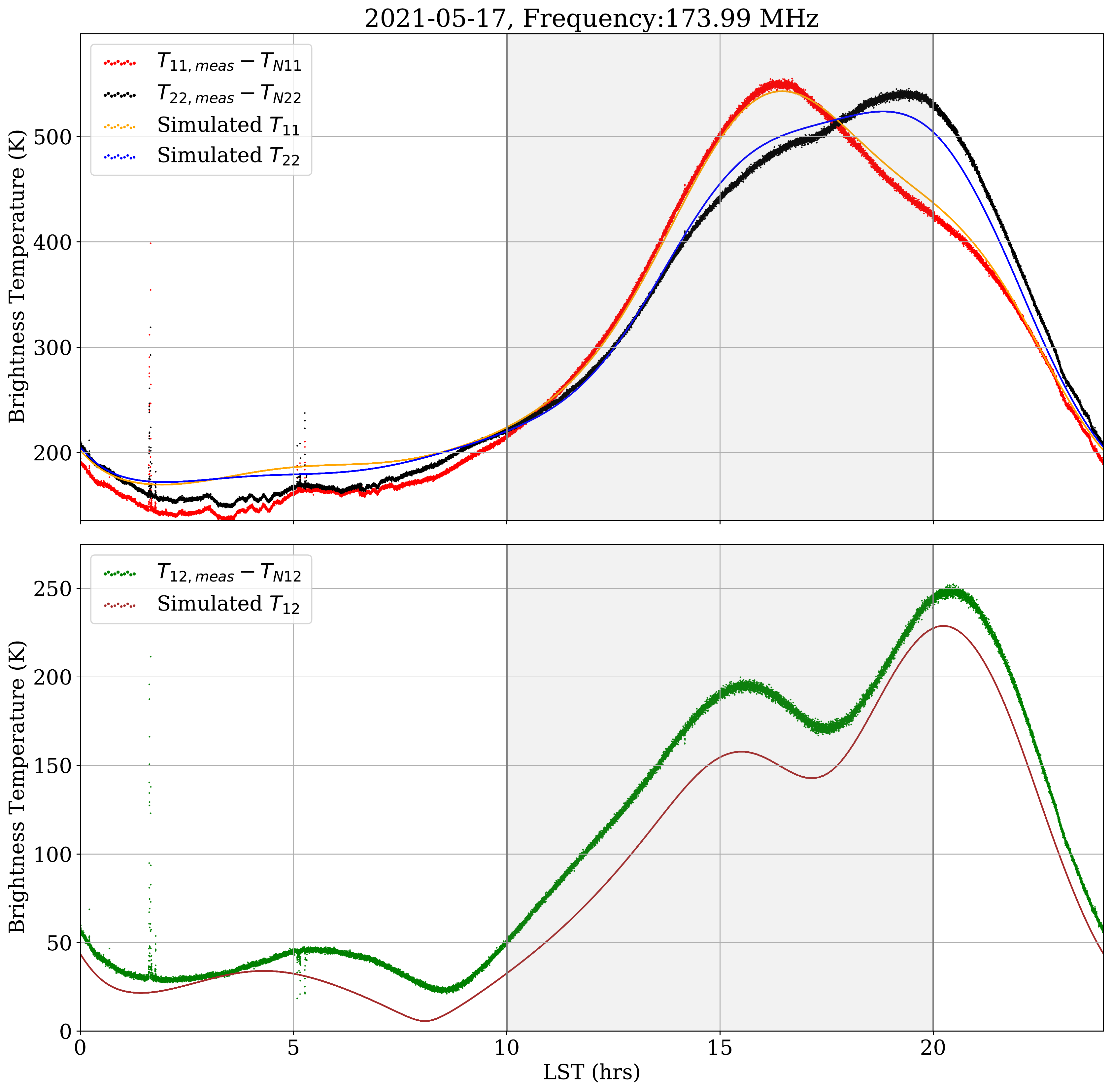}
\caption{Calibrated and $T_{Nij}$ subtracted data for $\sim 174$~MHz. The top panel shows calibrated auto-correlations along with simulated auto-correlations and the bottom panel shows calibrated cross-correlations along with simulated cross-correlations. The plot is of the same nature as Fig.\ref{fig:autocal_lst}, however at this frequency the individual antenna radiation patterns differ. Despite this being captured by the FEE simulations, the simulated temperatures differ from calibrated data.}
\label{fig:autocal_lst_170}
\end{figure*}

We now inspect the result of calibration at frequencies where the antenna radiation patterns differ substantially. Consider the plots in Fig.\ref{fig:autocal_lst_170} which are identical to those in  Fig.\ref{fig:autocal_lst} except that the frequency is now about 174~MHz. Despite scaling the temperatures as well as subtracting excess receiver noise temperatures, the shape of temperature $vs$ LST does not exactly follow the simulations, unlike the plots for 111~MHz. In inteferometers with closely spaced antennas, cross-talk between the antennas becomes non-negligible and has to be taken into account. Since SITARA has antennas spaced at 1~m, we attribute the differences between SITARA data and mock data to cross-talk between the antennas and attempt to model its effects. With a model including cross-talk, we expect to obtain better estimates of the receiver noise temperatures.

\subsection{An empirical model for cross-talk}
\label{subsec:cal_crosstalk}
In this section, we present an empirical model for the cross-talk in our system. We choose to model the data empirically such that the model can extended for future versions of SITARA with multiple antennas. The aim of this modelling is to enable \textit{forward modelling} of global 21~cm templates into the instrument plane and to search for them in the data after foreground subtraction etc. In \ref{Ap:cross_talk_phymodel}, we also present a plausible physical model for the terms in the empirical model.

If there were no cross-talk, individual antenna voltages would consist only of the signals induced on the specific antennas. In the presence of cross-talk, cross-correlations would have non-negligible amounts of auto-correlations and vice-versa. We may therefore model the resulting measurements as a combination of "ideal" auto-correlations and cross-correlations. If we ignore reflections, the problem can be linearised. Thus, the equations for auto and cross correlations in the presence of cross-talk at each frequency and LST can be written as
\begin{align}
    \label{Eq:cross_vis_eq}
    T_{11} = a_{11}V_{11} + a_{12}V_{12} + a_{21}V_{21} + a_{22}V_{22} + T_{n11} \\ \nonumber
    T_{12} = b_{11}V_{11} + b_{12}V_{12} + b_{21}V_{21} + b_{22}V_{22} + T_{n12} \\ \nonumber
    T_{21} = c_{11}V_{11} + c_{12}V_{12} + c_{21}V_{21} + c_{22}V_{22} + T_{n21} \\ \nonumber
    T_{22} = d_{11}V_{11} + d_{12}V_{12} + d_{21}V_{21} + d_{22}V_{22} + T_{n22}
\end{align}
where $V$ are the expected (simulated) visibilities in the absence of cross-talk, $T$ are the visibilities in the presence of cross-talk. Though $T_{21} = T_{12}^*$ and $V_{2,1} = V_{1,2}^*$, we have included them for the sake of completeness. For a drift scan instrument such as SITARA, $T$ and $V$ vary as a function of LST, while their coefficients are expected to be constant. We may thus write them as matrices as follows
\begin{equation}
\label{Eq:sys_matrix_emp} 
\bm{T} = \bm{VB},
\end{equation}
with $\bm{T}$, $\bm{V}$ and $\bm{B}$ given by Eqs.\ref{Eq:mat_V}, \ref{Eq:mat_B} and \ref{Eq:mat_T} respectively.
\begin{equation}
\label{Eq:mat_T}
\bm{T} = 
\begin{bmatrix}
T_{11}(t_i) & T_{12}(t_i) & T_{21}(t_i) & T_{22}(t_i) \\
\end{bmatrix}; i = 1~to~n\\
\end{equation} 
\begin{equation}
\label{Eq:mat_V}
\bm{V} = 
\begin{bmatrix}
V_{11}(t_i) & V_{12}(t_i) &  V_{21}(t_i) & V_{22}(t_i) & 1 \\
\end{bmatrix}; i = 1~to~n\\
\end{equation}
\begin{equation}
\label{Eq:mat_B}
\bm{B} = 
\begin{bmatrix}
a_{11} & b_{11} & c_{11} & d_{11}  \\
a_{12} & b_{12} & c_{12} & d_{12}  \\
a_{21} & b_{21} & c_{21} & d_{21}  \\
a_{22} & b_{22} & c_{22} & d_{22}  \\
T_{n11} & T_{n12} &T_{n21} & T_{n22}  \\
\end{bmatrix}
\end{equation}
where $\bm{T}$ is a $n~x~4$ matrix of measurements, $V$ is a $n~x~5$ matrix of simulated visibilities and $\bm{B}$ is a $5~x~4$ matrix of model coefficients.

If the measurements are not calibrated to brightness temperature scale i.e. if the measured and expected visibilities are not in the same units, we may write them as $\bm{P} = \bm{T}\bm{G}$ where $\bm{P}$ is a matrix with the measured, uncalibrated data. $\bm{G}$ is a $4x4$ complex diagonal gain matrix that has the signal chain gains as diagonal elements as given in Eq.\ref{Eq:matrix_G}. Being a diagonal matrix, the effect of $\bm{G}$ is just a scaling of the data.
\begin{multline}
\label{Eq:matrix_G}
G = 
\begin{bmatrix}
|G_1|^2 & 0 & 0 & 0 \\
0 & G_1G_2^* & 0 & 0 \\
0 & 0 & G_2G_1^* & 0 \\
0 & 0 & 0 & |G_2|^2 \\
\end{bmatrix} \\
\end{multline}
where $G_1$ and $G_2$ are the gains of the individual signal chains. This gives us the equation describing SITARA data as : 
\begin{equation}
\label{Eq:Sys_emp_matrix_comp_g}
\bm{P} = \bm{TG} = \bm{VBG}
\end{equation}
If $\bm{G}$ is accurately known, it can be inverted to calibrate our measurements $\bm{P}$ to units of kelvin, as $\bm{G}$ is generally non-singular and invertible. Though gains obtained in Sec.\ref{subsec:auto_cal} can be used to construct $\bm{G}$, it is also shown that the model used to obtain those gains is incomplete. Therefore, we will use raw measurements $\bm{P}$ in our subsequent analysis.

If we have $n>5$ independent measurements, it is possible to find a least-squares solution to Eq.\ref{Eq:Sys_emp_matrix_comp_g} to obtain the matrix of coefficients $\bm{BG}$. SITARA observations have a cadence of about 3 seconds and each observation spans several hours with good LST coverage and therefore, the $n>5$ condition is easily satisfied for $\bm{P}$. Also, visibilities simulated for each observational timestamp in the same fashion as in Sec.\ref{subsec:auto_cal} form $\bm{V}$, and for the same considerations given there, we use LSTs between 10 and 20 hours for $\bm{T}$ and $\bm{V}$. Eq.\ref{Eq:Sys_emp_matrix_comp_g} is then solved at each frequency using a least-squares algorithm (such as \texttt{numpy.linalg.lstsq}). It has to be noted that since we have not corrected data for $\bm{G}$, the solution that we obtain is $\bm{B}\bm{G}$ which is a product of coefficient matrix and gain matrix. In this work, we are interested in finding whether simple cross-talk considerations can better model the data.

Before proceeding to inspect the results of the least-squares fit, it is instructive to compare the above formalism with the procedure given in Sec.\ref{subsec:auto_cal}. It is easy to see that if cross-talk is neglected, the principal diagonal elements of $\bm{B}$ will have a value of unity, the last row will have values of the receiver noise temperatures and all other terms will be zero. Then, a least-squares solution provides an estimate of the gain matrix $\bm{G}$ as well as receiver noise temperatures that, as we have already noticed, are also less accurate. Therefore, it can be inferred that the procedure given in Sec.\ref{subsec:auto_cal} is a simplified form of the more generalised procedure given here. 

\begin{figure*}[hbt!]
    \centering
	\includegraphics[width=\columnwidth]{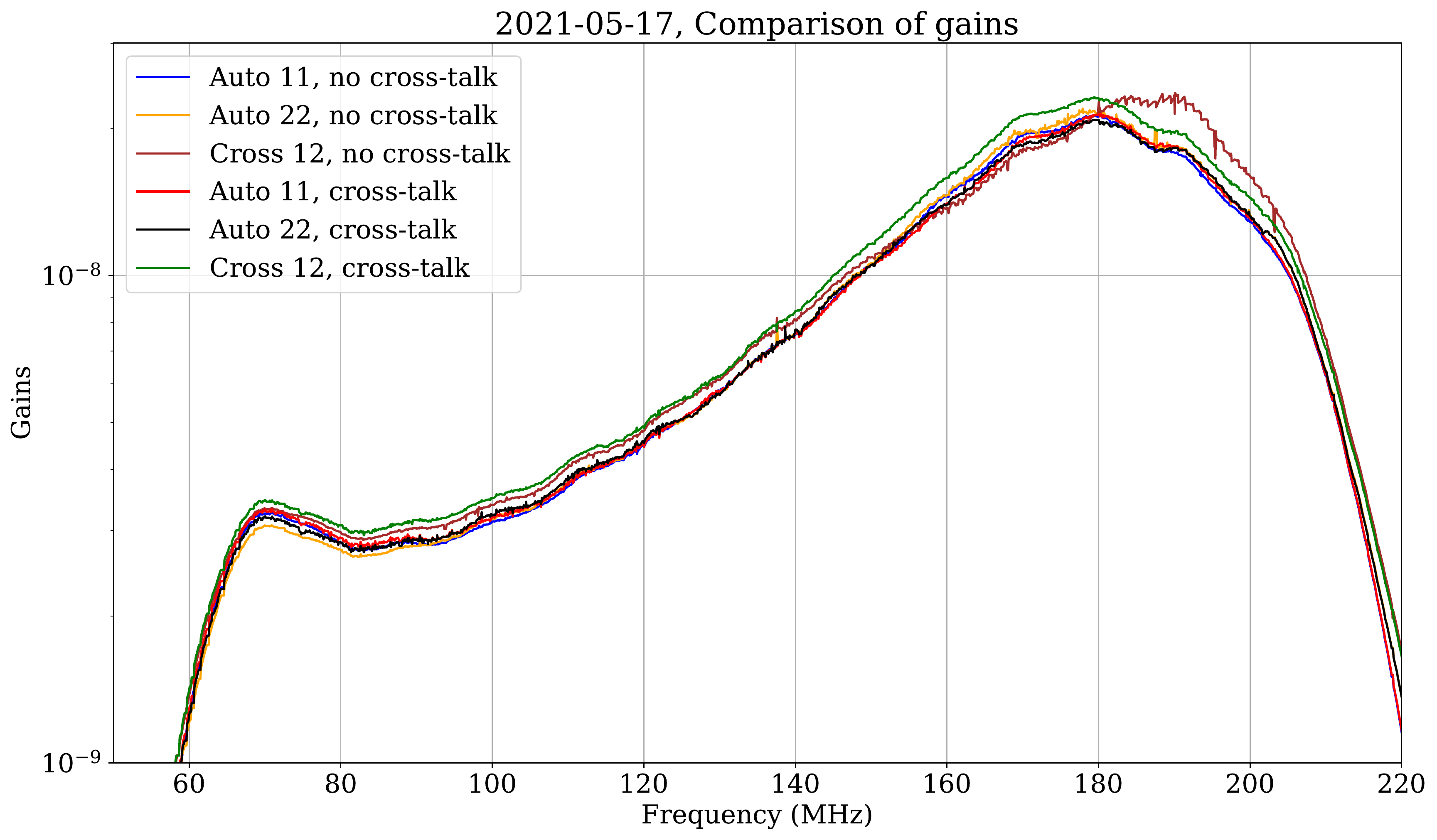}
    \caption{Comparison of gains estimated with and without cross-talk. The plots are semi-logarithmic to accommodate the dynamic range. As noted in the text, each "gain" in the cross-talk model is a sum of coefficients that includes cross-talk. Despite using two different formalism, it can be seen that they are in close agreement.}
    \label{fig:mat_g_sum}
\end{figure*}

\begin{figure*}[hbt!]
    \centering
	\includegraphics[width=\columnwidth]{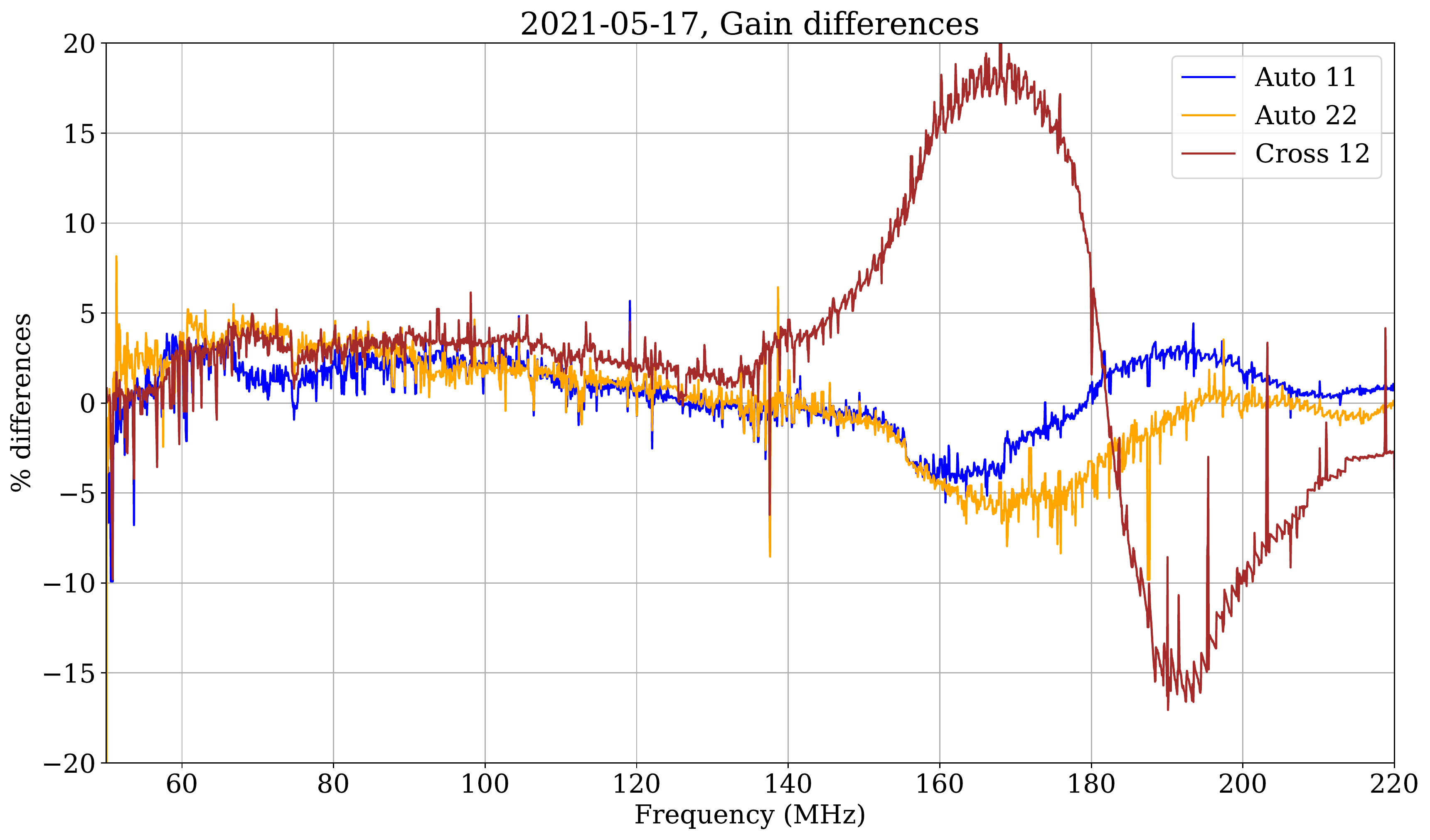}
    \caption{Differences between gains estimated with and without cross-talk. In this plot, the fractional differences between the gains estimated with and without cross-talk are shown as percentages. The auto-correlation gains derived with the two models have a difference less than 10\% while the cross-correlation gains differ by about 20\% at frequencies where the antennas patterns are dissimilar.}
    \label{fig:mat_g_diff}
\end{figure*}

We now inspect the resultant $\bm{BG}$ matrix. We have to note that due to the lack of accurate estimations of $\bm{G}$, "gain" becomes a concept which is not well defined in the cross-talk model. Besides, there could be linear dependencies in the matrix $\bm{V}$ that introduce degeneracies in the fitted parameters. For example, as the antenna patterns are nearly identical at frequencies less than 150~MHz, we expect $V_{11} \approx V_{22} \approx |V_{12}|$ and therefore their corresponding coefficients in matrix $\bm{BG}$ will be degenerate. Nonetheless, we expect a sum of the coefficients to mitigate such degeneracies. Therefore, to enable comparisons with the gains derived in Sec.\ref{subsec:auto_cal}, we use the sum of the coefficients in each column of $\bm{BG}$ (except the receiver noise temperatures) as a proxy for gains. Doing so also enables us to look at the differences between the two gain models. Fig.\ref{fig:mat_g_sum} compares the gains derived with and without cross-talk considerations and Fig.\ref{fig:mat_g_diff} shows the fractional difference as a percentage. We find that the differences are less than $10\%$ for auto-correlation gains and less than $20\%$ for cross-correlations, thus implying that the impact of cross-talk is less than $20\%$.

It is interesting to know whether the cross-talk model does a better job at accurately representing the measurements. While it is tempting to compute a pseudoinverse of $\bm{BG}$ and "calibrate" SITARA measurements $\bm{P}$, such an operation is erroneous. Therefore we choose to perform forward modelling to avoid issues of matrix inversion from affecting our results.
For this, simulated $\bm{V}$ and fitted $\bm{BG}$ are multiplied to generate mock SITARA data and compared with SITARA measurements $\bm{P}$. We also compare them with $\bm{V}$ modified by gains and receiver noise temperatures from the no cross-talk model (Sec.\ref{subsec:auto_cal}).
\begin{figure*}[hbt!]
    \centering
	\includegraphics[width=\columnwidth]{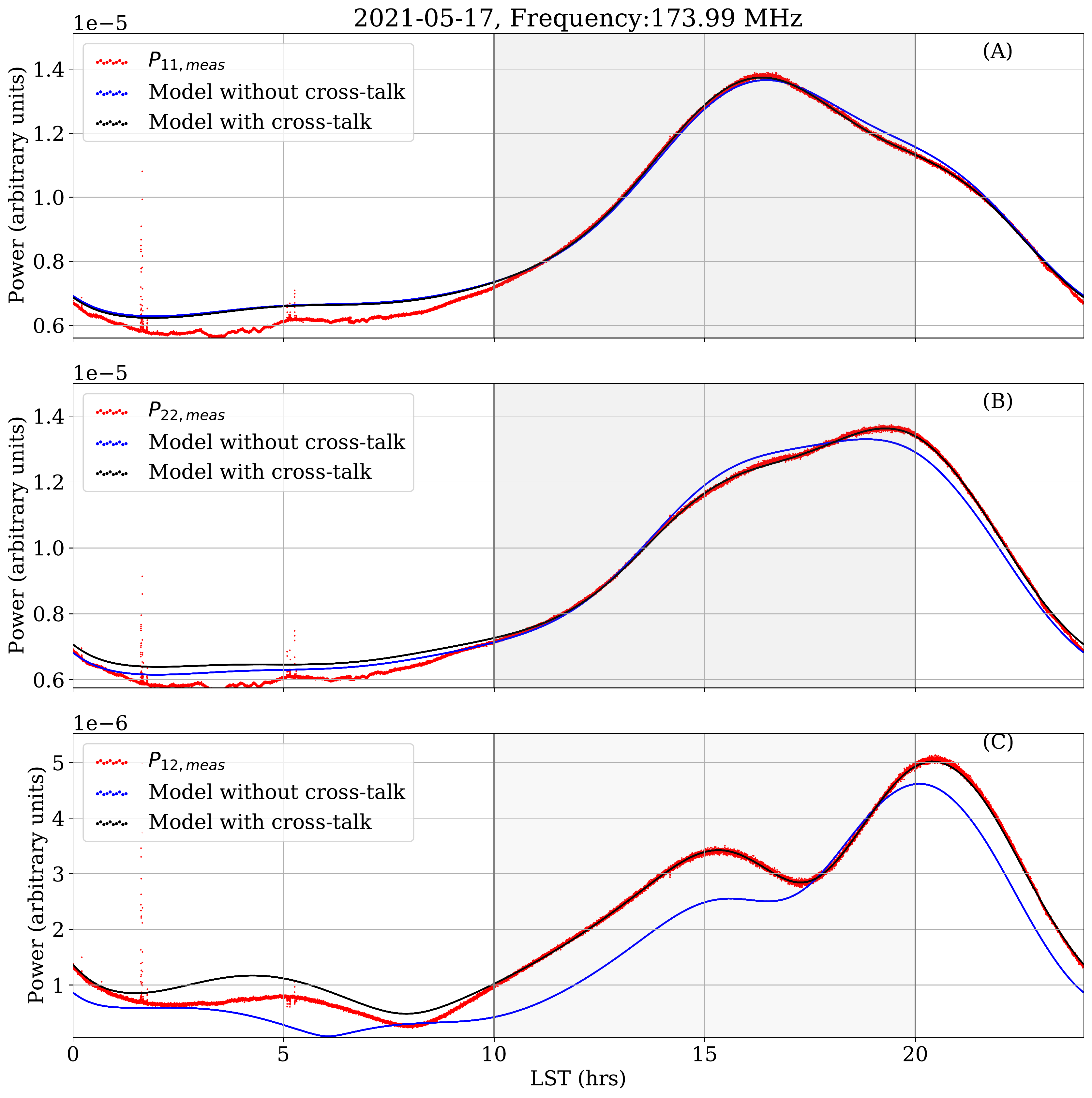}
    \caption{A comparison between SITARA data at 174~MHz with a model that does not consider cross-talk and one that considers cross-talk. Plots (A) and (B) are the auto-correlations and (C) is the cross-correlation. The data are forward modelled and therefore not in units of brightness temperature. Data from shaded area alone are used to compute gains and receiver noises. With the cross-talk model, the simulations match the data.}
    \label{fig:cross_talk_cal}
\end{figure*}

\begin{figure}[hbt!]
\centering
\includegraphics[width=\columnwidth]{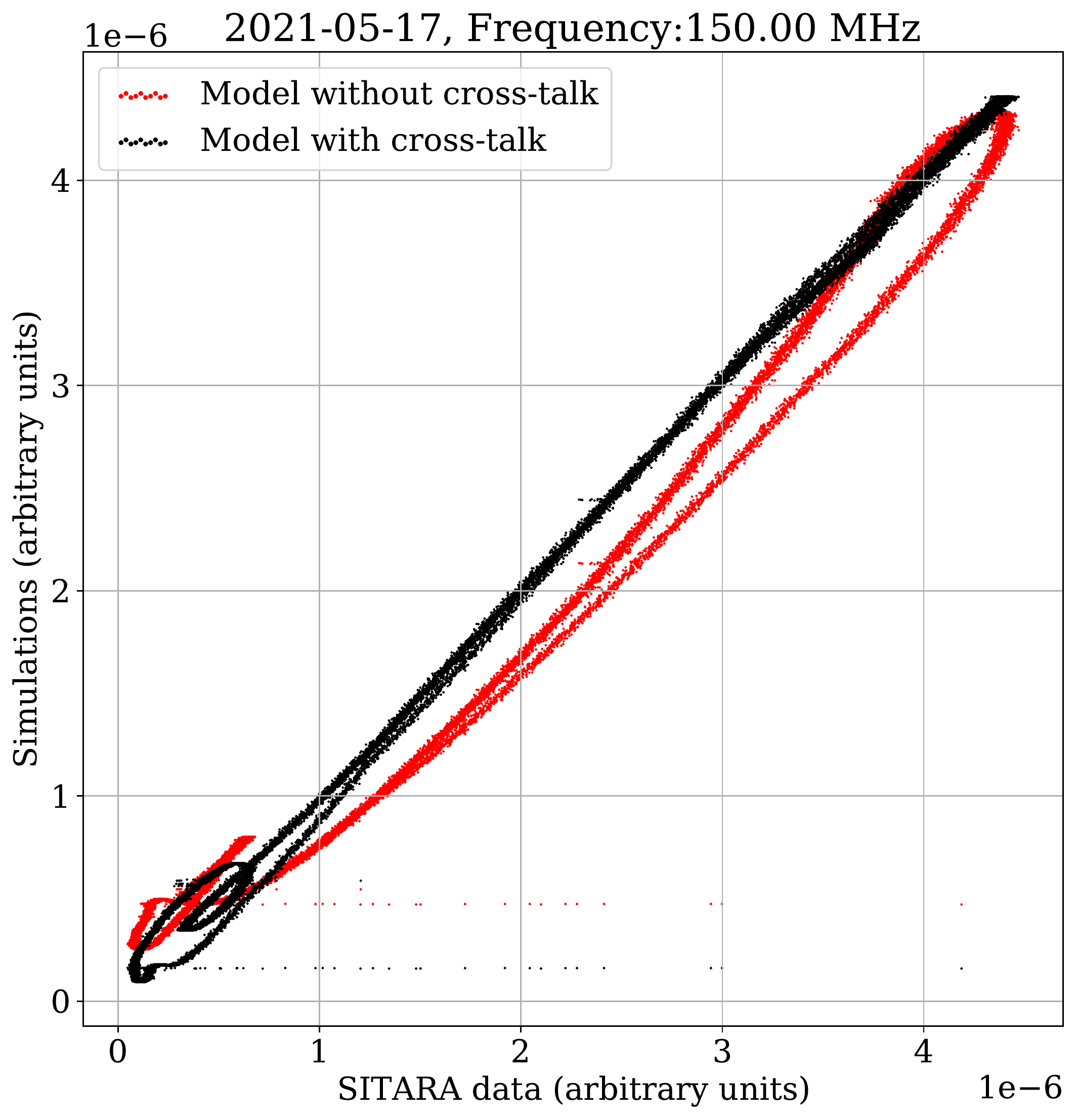}
\includegraphics[width=\columnwidth]{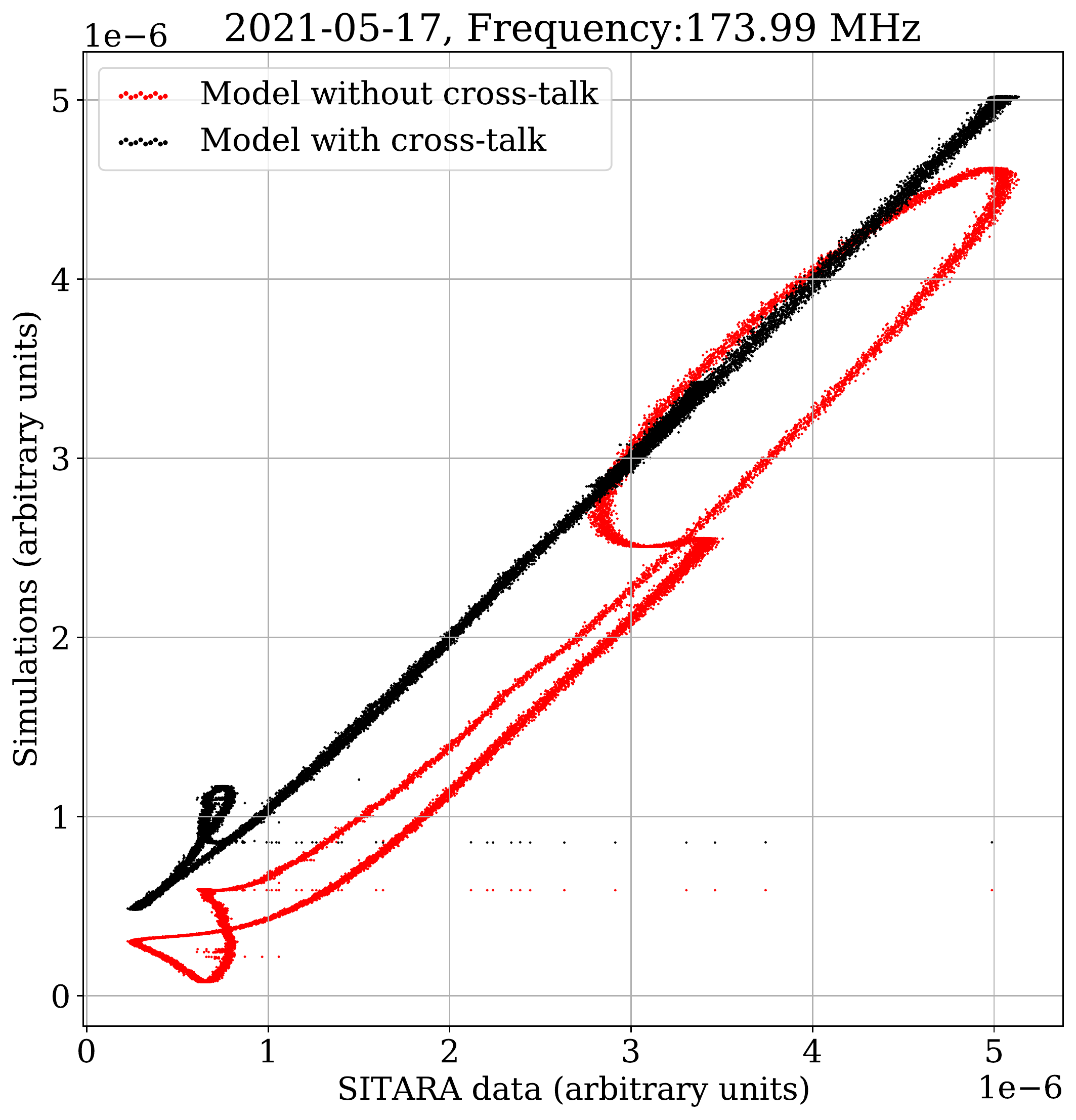}
\caption{Comparison between SITARA data and simulations for the cross-correlations. Shown are the temperature-temperature plots between the SITARA data and simulations based on the two models. Two frequencies where the individual antenna patterns are dissimilar are chosen. We expect the simulations to follow data in a linear fashion in this plot, if the model used for simulations is accurate. While the model neglecting cross-talk fails to explain the variations in data, the cross-talk model fits the data very well.}
\label{fig:cross_temp_temp_plots}
\end{figure}

Fig.\ref{fig:cross_talk_cal} shows the results of this forward modelling from which it is evident that the data are better modelled with a cross-talk model. A more informative way to represent the same data is to plot measurements against simulations using different models, as shown in Fig.\ref{fig:cross_temp_temp_plots}. Such a temperature-temperature plot would be a straight line if the data are well represented by the simulations. Once again, we find that the cross-talk model represents data better. 

\section{Results}
In this section, we present some of the results obtained from the data analysis given in previous sections. Specifically, we present measurements of coupled receiver noise in SITARA as well as attempt to answer the question of whether SITARA is sensitive to all sky signals.
\subsection{Coupled receiver noise}
\begin{figure*}[hbt!]
    \centering
	\includegraphics[width=\columnwidth]{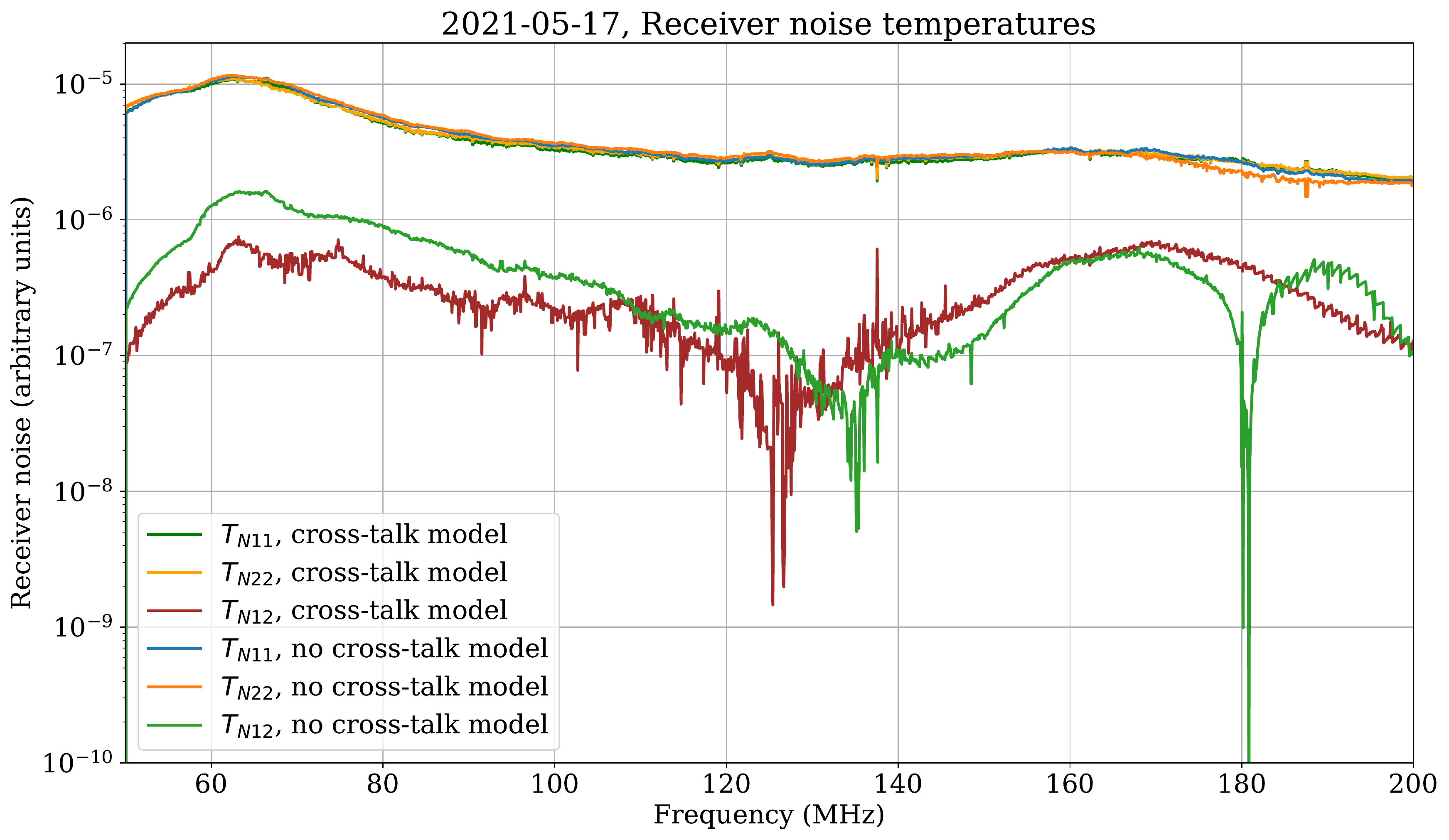}
    \caption{A comparison between estimations of receiver noise with and without cross-talk considerations. The receiver noise estimates are not calibrated to units of kelvin. It is seen that the estimations of coupled receiver noise are generally lower when cross-talk is modelled.}
    \label{fig:cross_noise}
\end{figure*}

As a by-product of the modelling performed in Sec.\ref{subsec:cal_crosstalk} that includes cross-talk, we obtain estimates of the receiver noise temperatures $T_{n11}, T_{n22}$ and $T_{n12}$. However, these receiver noises are not calibrated to units of kelvin, and therefore have an arbitrary scaling introduced by the instrumental gains. Thus, we refer to them as $P_{n11}, P_{n22}$ and $P_{n12}$. In a similar vein, we forward model the receiver noise temperatures obtained in Sec.\ref{subsec:auto_cal} where cross-talk has been ignored, to enable a comparison between the two methods, shown in Fig.\ref{fig:cross_noise}.

\begin{figure*}[hbt!]
    \centering
	\includegraphics[width=\columnwidth]{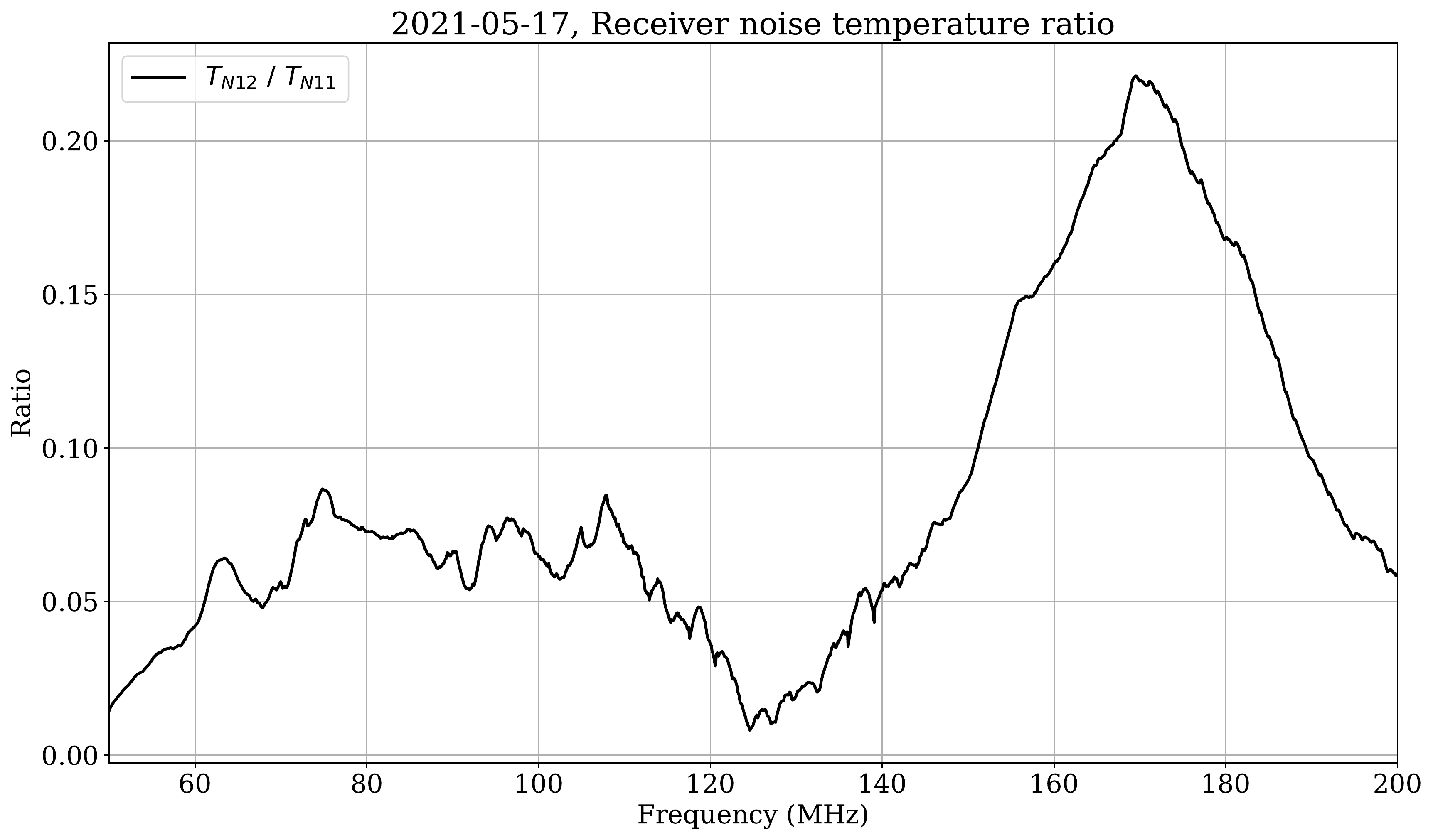}
    \caption{Ratio of estimated coupled receiver noise temperature to an auto-correlation receiver noise temperature. As expected, the cross-coupled receiver noise in data is substantially lower than auto-correlation receiver noise. The data have been smoothed with a Savitzky-Golay filter to reduce noise in the plots.}
    \label{fig:cross_noise_ratio}
\end{figure*}

We find that the coupled receiver noise estimations with a model containing cross-talk are lower than the ones estimated without cross-talk. However, Fig.\ref{fig:cross_noise} shows data that are not in units of temperature as the gains are not accurately known, thereby limiting its utility. We therefore use a ratio of the coupled and self noises as shown in Fig.\ref{fig:cross_noise_ratio} to mitigate the effects of gain. It is seen that the cross-coupled receiver noise is approximately 10\% of the receiver noise in auto-correlations; which has some interesting consequences. As discussed in Sec.\ref{subsec:mean_sky_sens}, interferometers have sensitivity to a uniform component that is a function of frequency. Frequencies less than 150~MHz where SITARA has more than 10\% sensitivity to uniform components also happen to be where the ratio of coupled to self noise is less than 5\%. While a subtraction of coupled receiver noise requires detailed modelling of the receiver noise coupling between interferometer arms (see for e.g. \cite{sutinjo_ursi}), their reduced levels by an order of magnitude compared to the receiver temperatures in auto-correlations may reduce their deleterious impact in detecting 21~cm signals. 

\subsection{Is SITARA sensitive to an all-sky signal ?}
\label{subsec:mean_sky_sens}
A global component of the sky is that average of the sky temperature over the entire solid angle visible to an observer. While previous works have demonstrated that an interferometer does respond to a uniform component by numerically integrating Eq.\ref{Eq:coh_1} for various antenna types and orientations, no experiment has conclusively demonstrated these simulations. In this section, we attempt to answer the question of whether an interferometer becomes sensitive to an all sky signal at short baselines using broadband data collected with SITARA.

In order to enable comparison between measured data and simulations such as in \cite{Singh_2015}, we define a quantity that is called coherence in Eq.\ref{eq:Coherence_2}. The ratio of the measured auto-correlations and cross correlations provides an estimate of the coherence that an interferometer would measure at short baselines. In Eq.\ref{eq:Coherence_2}, the coherence provided is devoid of the individual receiver gains. Owing to a formal resemblance between this computed quantity and the "complex degree of coherence" in optics \citep{1959pot1.book.....B}, we refer to this quantity as the degree of coherence. We use the letter $C$ to denote the degree of coherence instead of $\Gamma$ used in optics, as $\Gamma$ stands for reflection coefficient in electromagnetics.
\begin{align}
\label{eq:Coherence_2}
C(\nu) &= \frac{T_{12}(\nu)}{\sqrt{T_{11}(\nu) T_{22}(\nu)}} \nonumber \\
       & \approx \frac{P_{12}(\nu)}{\sqrt{P_{11}(\nu) P_{22}(\nu)}}
\end{align}
where $T_{12}(\nu)$ is the measured cross-correlations, $T_{11}(\nu)$ and $T_{22}(\nu)$ are corresponding auto-correlations, all with receiver noise  subtracted. Since coherence is a ratio of temperatures, we expect it to be independent of instrument gains, and a ratio of raw powers can be used instead (with receiver noise subtracted). For a uniform sky of unit temperature $T_{A11} = T_{A22} = 1$, coherence is simply the visibility in cross-correlations. As shown in \cite{Singh_2015}, the visibility as a function of antenna spacing (or equivalently as a function of frequency for a fixed physical baseline) is expected to have a characteristic $sinc$ shape for a uniform sky temperature. Viewed in this light, the simulated cross-correlations for a uniform sky can be interpreted as the degrees of coherence for the same. 

The simplified picture given above is complicated by the presence of foregrounds having spatial structures, antennas with anisotropic radiation patterns, antenna pattern variations as a function of frequency (the so called "chromaticity" which couples spatial structures into the measured visibilities), noise coupling and cross-talk between the antennas and signal reflections within the receiver chains. Of these, foregrounds have the largest impact on our measurements.
The low frequency radio sky above $\sim$ 10~MHz is dominated by Galactic synchrotron emission largely following a power law spectrum that also has spatial variations in the spectral index \citep{10.1093/mnras/sty3410, Rao_2016}. If the foregrounds were spatially uniform, it is easy to see that their effect on coherence would be similar to that of a uniform sky. However, foregrounds have substantial structure, especially on the Galactic plane, and the measured visibilities would be the antenna beam weighted sum of these structures. 

Thus, it is imperative to use sky regions with minimal structures - that can be considered close to a uniform sky -  to carry out coherence computations. We expect imaginary components of visibilities to be minimal when the sky has no substantial off-zenith structures. If a minimum in imaginary is caused by bright compact sky regions at zenith - which is the nominal phase centre of SITARA - such a minimum would be short lived as the compact regions drift away from the zenith. Therefore, we choose a subset of data with the lowest imaginary component across all frequencies in cross-correlations, which also have the least variation in imaginary components as a function of time. We find LSTs between 4 to 5 hours to satisfy these conditions. 

\begin{figure*}[hbt!]
    \centering
	\includegraphics[width=\columnwidth]{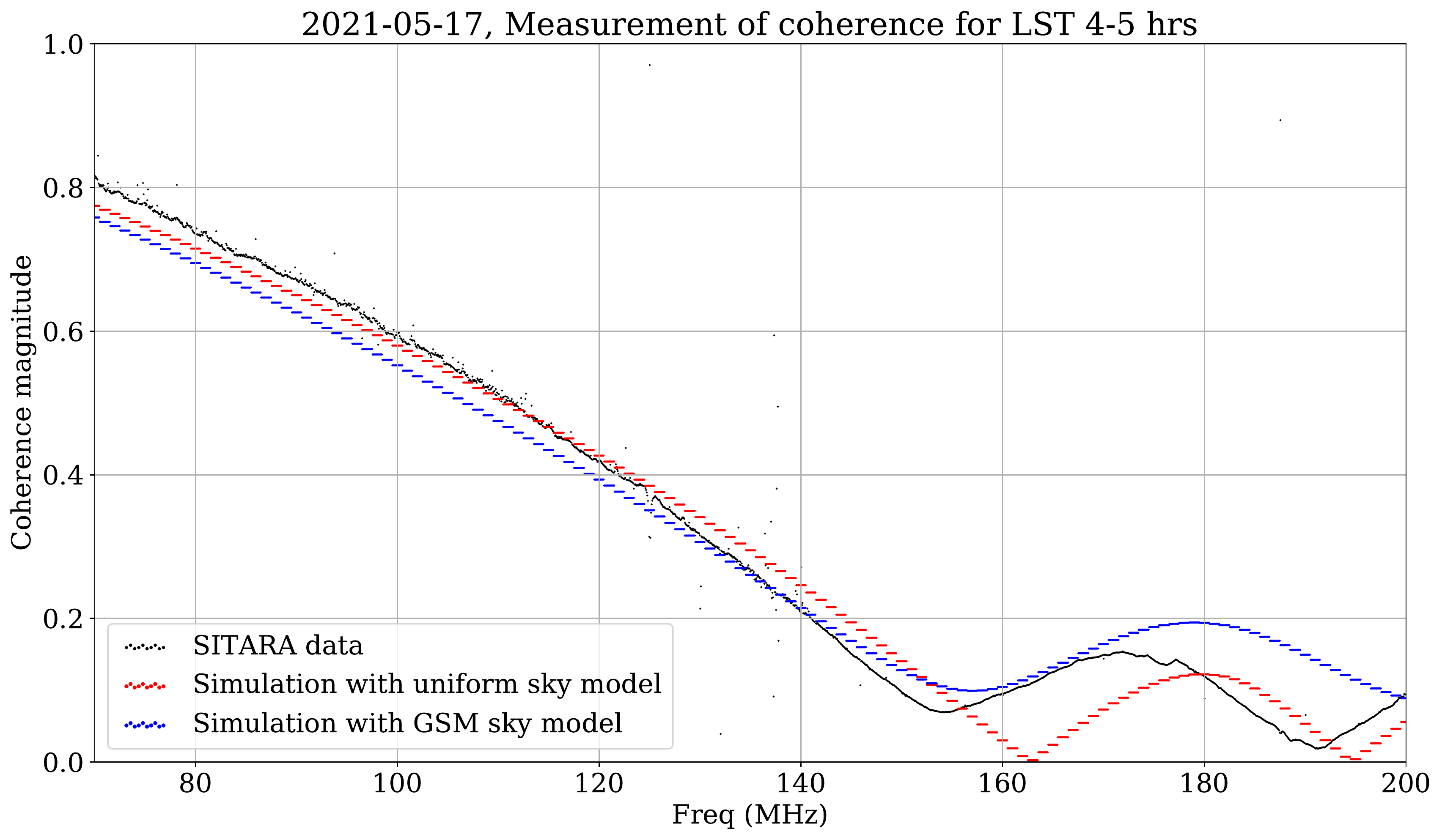}
    \caption{Comparing measured coherence (black) with simulations assuming a uniform sky (red) and a more realistic GSM foregrounds (blue). Uncalibrated data with receiver noise subtracted from auto and cross-correlations are used for this computation. }
    \label{fig:Coher_plot}
\end{figure*}

Since LSTs of 4-5 hours for the data used in this work corresponded to local day time when we expect the system gains to be different to the ones from night time, we estimate $\bm{BG}$ matrix with all 24 hours of data. From $\bm{BG}$ we obtain estimates of receiver noise temperatures $P_{n11}, P_{n22}$ and $P_{n12}$ and subtract them from the respective averaged auto and cross correlations. However, we do not remove other cross-talk components from the calibrated data to keep the computations simple. The coherence, as a function of frequency as given in Eq.\ref{eq:Coherence_2}, is then computed, and the results are given in Fig.\ref{fig:Coher_plot}. The same figure also shows simulated coherence employing the method given in \ref{subsec:auto_cal}, with a uniform sky model as well as a more realistic GSM foreground model. 

The close similarity between the measured coherence and the simulated ones, despite neglecting the cross-talk in the measurements, shows that SITARA is indeed sensitive to an all-sky component. 

\section{Discussion}
Based on our modelling and analysis of SITARA data, we draw the following inferences.
\begin{enumerate}
    \item Interferometers with short baselines are sensitive to all sky signals such as uniform components, with a response that closely matches simulations following \cite{Presley_2015} and \cite{Singh_2015}. 
    \item Such interferometers also have non-negligible cross-talk as well as internal noise coupling between the antennas that has to be considered when they are employed for precision cosmology. 
\end{enumerate}
It is interesting to compare Fig.\ref{fig:cross_noise_ratio} and Fig.\ref{fig:Coher_plot} with theoretical predictions of \cite{Venumadhav_2016}, specifically their Fig.5 that predicts that the spectral shapes of coherence and noise have an anti-correlation. Despite the differences between the SITARA setup and the case studied in \cite{Venumadhav_2016}, we find that the trend followed by the spectral shapes reported in this work are identical to the ones shown in \cite{Venumadhav_2016}. It is seen that at frequencies larger than 150~MHz where the coherence is low, the coupled receiver peaks. However, this observation has to be treated with some caution as \cite{Venumadhav_2016} consider electrically short dipoles while SITARA antennas are not electrically short at all frequencies.

The main utility of this work is in short spacing interferometry in a 21~cm context. Nonetheless, the systematic effects seen in SITARA are expected to be seen by interferometers with closely packed antennas such as EDA-2 and SKA-low \citep{SKA_Low_report, 2020SPIE11445E..89V} stations. Extensions of the work given in this paper may also aid in the analysis of systematics in those instruments as well as aid in their calibration. In this context, we also would like to point out that some of the cm-wave CMB instruments with closely packed elements observed excess spurious contributions in the data that were never fully explained \citep{10.1046/j.1365-8711.2003.06338.x, Padin_2001}. While the optics and electronics of the cm-wave instruments differ from low-frequency instruments such as SITARA, analysis based on adaptions of the empirical cross-talk model that we have outlined may aid in understanding the systematics in such instruments better.  
\subsection{Future work}
We have not provided error estimates in this work. While it is possible to provide \textit{formal} fitting errors based on the covariances of the fits performed, we have found them to be less reliable and frequently underestimating the errors. In future work, we expect to present an extensive error model devoid of such issues. A caveat with the approach given in this work is that the calibration and temperature scales are tied to diffuse sky models provided by the GSM. Many of the maps used as inputs to the GSM themselves have errors that are poorly understood, and re-calibrations of these maps are required when compared with precision radiometric data (see for e.g. \cite{Patra_2015}). Noise-source based bandpass calibration of signal chains can be employed, along with in-situ measurements of complex antenna scattering parameters, to enable better modelling of various effects such as cross-talk. Use of compact, integrated circuitry deployed in active antennas such as LEDA \citep{doi:10.1093/mnras/sty1244} can provide stable (with careful designs, traceable) calibration states to perform high time cadence calibration. However, since SITARA is an interferometer, techniques to provide \textit{correlated} noise to the antennas have to be explored. 

A potential source of error in low frequency radiometric calibration is the antenna pattern model. Since SITARA relies on simulated radiation patterns and sky models to calibrate the instrument, errors in either lead to mis-calibration. In-situ measurements of antenna patterns with unmanned aerial vehicles (UAVs) equipped with radio frequency instrumentation  \citep{9135792} or satellites \citep{Chokshi2020} can aid in this aspect. 

Alternative calibration strategies include using multiple antennas and closure relations. The techniques developed in this paper are currently being attempted with sky drift data acquired with EDA-2 at select frequencies (McKinley et.al. in prep). We also plan to carry out observations with different antenna spacings and orientations to quantify the response of interferometers at various spacings. 

\section{Conclusions}
In this paper we have detailed the system design, development and deployment of a short spacing interferometer - SITARA. We have also described the calibration strategies and some initial results employing those strategies. We find that interferometers with short baselines do have a response to all-sky signals. We also find that they have non-negligible cross-talk as well as noise coupling, with noise coupling less than 20\% of the individual receiver noises in the current configuration of SITARA. We plan to modify the SITARA system and carry out observations as well as evolve the techniques so as to make them useful for similar closely packed interferometers such as EDA-2 and SKA-low.

\section*{Acknowledgements}
This  work  is  funded  through CT's  ARC  Future Fellowship, FT180100321 - "Unveiling the first billion years:  enabling Epoch of Reionisation science". The International Centre for Radio Astronomy Research (ICRAR) is a Joint Venture of Curtin University and The University of Western Australia, funded by the Western Australian State government. This scientific work makes use of the Murchison Radio-astronomy Observatory, operated by CSIRO. We acknowledge the Wajarri Yamatji people as the traditional owners of the Observatory site. SITARA is an external instrument at the MRO, and is supported by MWA infrastructure at the MRO. Support for the operation of the MWA is provided by the Australian Government (NCRIS), under a contract to Curtin University administered by Astronomy Australia Limited. We acknowledge the Pawsey Supercomputing Centre which is supported by the Western Australian and Australian Governments.

We would like to acknowledge the expertise of the following people who contributed to the development of SITARA. David Kenney and Clinton Ward contributed to the hardware development and EMI chamber measurements of SITARA. A test deployment of SITARA to validate its end-to-end performance was carried out at John Kennewell's Australian Space Academy \footnote{\url{https://spaceacademy.net.au}}, Meckering, WA with the help of John Kennewell and Randall Wayth. The MWA operations team at ICRAR/Curtin assisted in the deployment of SITARA; specifically Andrew McPhail and David Emrich played a crucial role in SITARA deployment at the MRO and troubleshooting.  Finally, we would like to thank the anonymous reviewer whose valuable insights have helped in improving the quality of this paper. 

This work uses the following python packages and we would like to thank the authors and maintainers of these packages :- \texttt{numpy }\citep{2020Natur.585..357H}, \texttt{scipy }\citep{2020NatMe..17..261V}, \texttt{healpy }\citep{2019JOSS....4.1298Z}, \texttt{astropy }\citep{astropy:2013, astropy:2018}, \texttt{aipy }\citep{2016ascl.soft09012P}, \texttt{matplotlib }\citep{Hunter:2007}, \texttt{ephem }\citep{2011ascl.soft12014R} and \texttt{pygdsm }\citep{2016ascl.soft03013P}.

\section*{Data availability}
The data used in this work will be made available based on reasonable requests.


\bibliography{SITARA_pasa}

\appendix

\section{A physical model for cross-talk}
\label{Ap:cross_talk_phymodel}
In Sec.\ref{subsec:cal_crosstalk}, an empirical model for cross-talk is provided. While the empirical model does describe the data, a model that is physically motivated would help in system design and analysis. Here we provide a plausible physical model, though we do not fit this model to our data. We would like to emphasise that the model presented here is a very simplistic one; in reality it is \textit{not} possible to parameterise the cross-talk into a single factor as the cross-talk depends on the LNA input impedance and noise parameters. A more detailed model will be explored in future work.

We parameterise the cross-talk by a factor $f_c$ which is the fraction of voltage that gets coupled from one antenna (or any where along its signal chain) to the other.  We assume the two-antenna system to be reciprocal and hence $f_c$ is same for both antenna~1-2 and antenna~2-1 paths. 
Under these assumptions, we may write the voltages at the antenna terminals as 
\begin{align}
    \label{Eq:coupled_voltages_1}
    e_{1}(\nu, LST) = e_{1, sky}(\nu, LST) + e_{1, RX}(\nu) + f_c(\nu) e_{2}(\nu, LST) \\ \nonumber
    e_{2}(\nu, LST) = e_{2, sky}(\nu, LST) + e_{2, RX}(\nu) + f_c(\nu) e_{1}(\nu, LST)
\end{align}
where $e_{n, sky}$ are the voltages induced by external radiation (sky) on the individual antennas and $e_{n, RX}$ are the internal receiver noise voltages. However, Eqs.\ref{Eq:coupled_voltages_1} are coupled to each other and pose challenges in their application. Therefore we ignore cross-talk that arises from multiple couplings back and forth between the antennas, leading to Eqs.\ref{Eq:coupled_voltages_2}.
\begin{align}
    \label{Eq:coupled_voltages_2}
    e_{1} = e_{1, sky} + e_{1, RX} + f_c [e_{2, sky} + e_{2, RX}] \\ \nonumber
    e_{2} = e_{2, sky} + e_{2, RX} + f_c [e_{1, sky} + e_{1, RX}]
\end{align}
We can now form auto-correlations and cross-correlations from these voltages as $T_{ij} = e_i e_j^*$. 
\begin{equation*}
    \label{Eq:coupled_visi_a11}
\begin{split}
    T_{11} = |e_{1, sky}|^2 + e_{1, sky}e_{2, sky}^*f_c^* + e_{1, sky}^*e_{2, sky}f_c + |f_c|^2|e_{2, sky}|^2 \\
    + |e_{1, RX}|^2 + |f_c|^2|e_{2, RX}|^2
\end{split}
\end{equation*}
\begin{equation*}
\label{Eq:coupled_visi_a22}
\begin{split}
    T_{22} = |e_{2, sky}|^2 + e_{2, sky}e_{1, sky}^*f_c^* + e_{2, sky}^*e_{1, sky}f_c + |f_c|^2|e_{1, sky}|^2 \\ 
    + |e_{2, RX}|^2 + |f_c|^2|e_{1, RX}|^2
\end{split}
\end{equation*}
\begin{equation}
\label{Eq:coupled_visi_c12}
\begin{split}
    T_{12} = f_c^*|e_{1, sky}|^2 + e_{1, sky}e_{2, sky}^* + |f_c|^2e_{2, sky}e_{1, sky}^* + f_c|e_{2, sky}|^2 \\ 
    + f_c^*|e_{1, RX}|^2 + f_c|e_{2, RX}|^2 
\end{split}
\end{equation}
Eqs.\ref{Eq:coupled_visi_c12} may be rewritten into a matrix form as given in Eq.\ref{Eq:Sys_matrix}, identifying $e_{i, sky}e_{j, sky}^* = V_{i,j}$ where $V_{i,j}$ are expected visibilities in the absence of cross-talk, as computed using Eq.\ref{Eq:Visi_1}. Also, we use $T_{n11} = |e_{1, RX}|^2 + |f_c|^2|e_{2, RX}|^2$, $T_{n22}=|e_{2, RX}|^2 + |f_c|^2|e_{1, RX}|^2$ and $T_{n12} = f_c^*|e_{1, RX}|^2 + f_c|e_{2, RX}|^2$ to denote the noise temperatures of instrumental origin. For a drift instrument such as SITARA, expected visibilities as well as measured correlations change as a function of LST, owing to the movement of various sky regions through the antenna beams. This has also been incorporated into Eq.\ref{Eq:Sys_matrix} as $1....n$ rows in the expected visibilities as well as data matrices. 
\begin{multline}
\label{Eq:Sys_matrix}
\bm{T} = \bm{VF}, \rm{where}  \\ 
\bm{T} = 
\begin{bmatrix}
T_{11}(t_i) & T_{12}(t_i) &  T_{21}(t_i) & T_{22}(t_i) \\
\end{bmatrix} ; i = 1~to~n\\
\bm{V} = 
\begin{bmatrix}
V_{11}(t_i) & V_{12}(t_i) & V_{21}(t_i) & V_{22}(t_i) & 1\\
\end{bmatrix} ; i = 1~to~n\\
\bm{F} = 
\begin{bmatrix}
1       & f_c^*   & f_c     & |f_c|^2 \\
f_c^*   & 1       & |f_c|^2 & f_c \\
f_c     & |f_c|^2 & 1       & f_c^*\\
|f_c|^2 & f_c     & f_c^*   & 1\\
T_{n11} & T_{n12} & T_{n21} & T_{n22} \\
\end{bmatrix} \\
\end{multline} 
$\bm{T}$ is the matrix of measured auto and cross correlations, $\bm{V}$ is the matrix of expected visibilities in the absence of any cross-talk and $\bm{F}$ is the matrix with the coefficients.
Given a set of simulated visibilities $\bm{V}$ and a set of measurements $\bm{T}$ in the same units as visibilities, i.e. kelvins, Eq.\ref{Eq:Sys_matrix} may be solved to obtain the matrix of coefficients $\bm{F}$. If the data are not calibrated to units of kelvins, the measurements will be treated as raw powers $\bm{P} = \bm{TG}$ where $\bm{G}$ is the gain matrix given in \ref{Eq:gain_matrix}.
\begin{multline}
\label{Eq:gain_matrix}
G = 
\begin{bmatrix}
|G_1|^2 & 0        & 0          & 0 \\
0       & G_1G_2^* & 0          & 0 \\
0       & 0        & G_1^*G_2   & 0 \\
0       & 0        & 0          & |G_2|^2 \\
\end{bmatrix} \\
\end{multline}

\begin{figure*}[hbt!]
    \centering
    \includegraphics[width=\columnwidth]{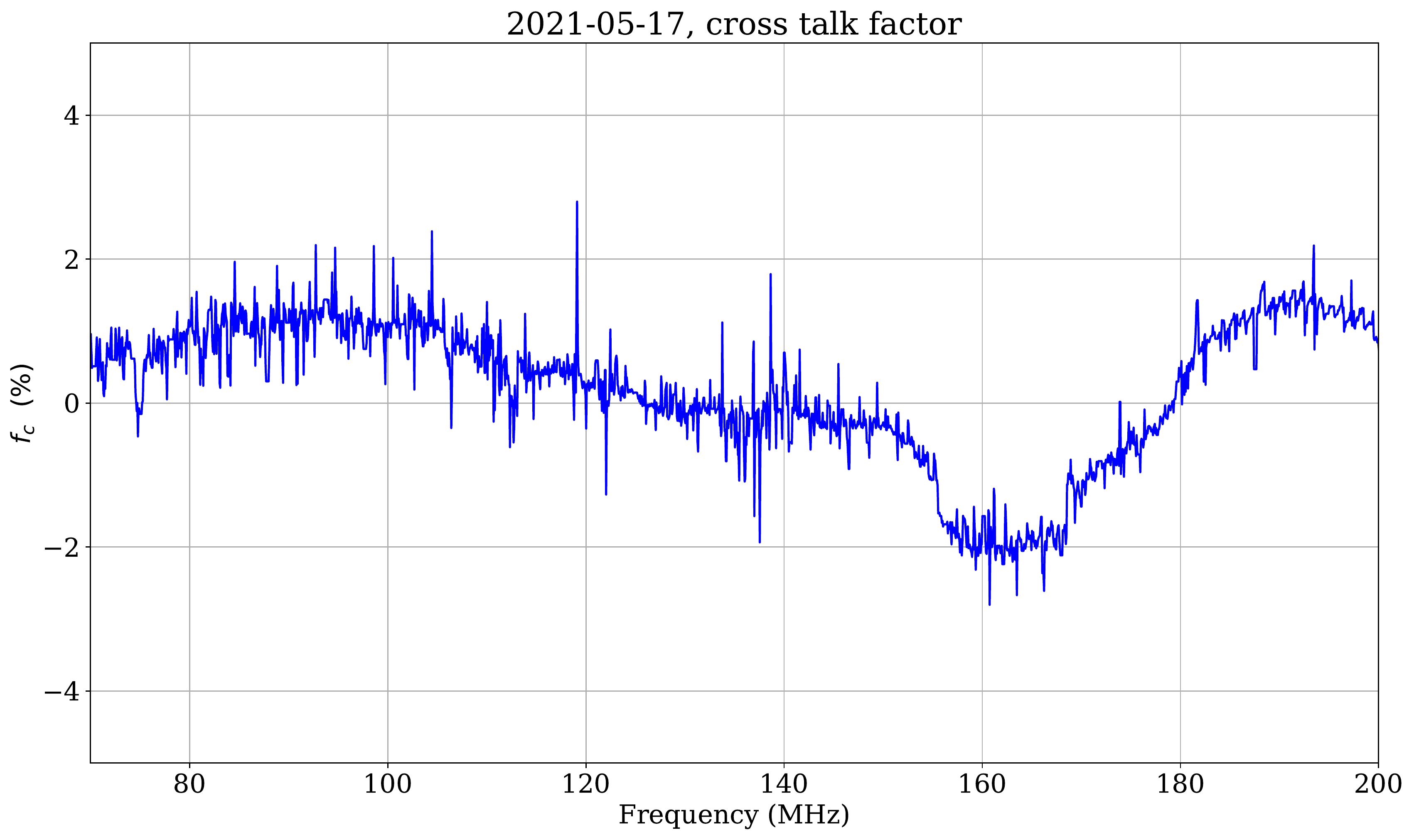}
    \caption{Cross-talk factor $f_c$ as a percentage. Please see the text for caveats associated with this calculation.}
    \label{fig:f_c_plot}
\end{figure*}

Comparing Eq.\ref{Eq:Sys_emp_matrix_comp_g} with the above formalism, it can be seen that $\bm{B}$ assumes the role of matrix $\bm{F}$ in empirical model, with the cross-talk coefficient $f_c$ represented by the coefficients such as $a_{21}, b_{22}, c_{11}, d_{12}$ etc. However, the unconstrained least-squares fitting used in the empirical model (Sec.\ref{subsec:cal_crosstalk}) does not preserve the relations between the coefficients. This, coupled to the fact that the gain matrix $\bm{G}$ is not accurately determined, makes establishing a relation between $\bm{B}$ and $\bm{F}$ a difficult exercise. However, if we assume that the auto-correlation based gains obtained in Sec.\ref{subsec:auto_cal} are reliable, it is possible to obtain some insights into the cross-talk in SITARA. For this, we sum the individual columns (except the receiver noise row) in $\bm{FG}$ into a factor $y_c$. For antenna-1 autocorrelations, this gives us 
\begin{align}
    \label{Eq:fac_y}
    y_{c1} &= |G_1|^2 \big(1 + f_c + f_c^* + |f_c|^2\big) \\ \nonumber
           &= |G_1|^2 \big(1+f_c\big) \big(1+f_c\big)^*
\end{align}
We further assume that the cross-talk factor $f_c$ is real valued, and  write 
\begin{align}
    \label{Eq:fac_y_real}
    y_{c1} &= |G_1|^2 \big(1 + 2f_c + f_c^2 \big)
\end{align}
Dividing $y_{c1}$ with the gain $|G_1|^2$ and taking the roots of the resulting quadratic equation, we obtain an estimate of $f_c$, which is shown in Fig.\ref{fig:f_c_plot} as a percentage. It has to be noted that there are caveats associated with this calculation. The assumptions that we made to obtain this estimate are not fully justified; calculation of $|G_1|^2$ is shown to be inaccurate and $f_c$ cannot be real valued at all frequencies due to the finite path lengths for cross-talk. Therefore, the calculated $f_c$ is given only to demonstrate an application of the physical cross-talk model.

\end{document}